\documentclass[12pt]{article}
\usepackage{tikz}
\usetikzlibrary{matrix,arrows}
\usepackage{jheppub}
\usepackage{amsmath,amssymb,euscript,array,mathrsfs}
\usepackage{arydshln}
\usepackage{todonotes}

\newcommand{\startappendix}{
\setcounter{section}{0}
\renewcommand{\thesection}{\Alph{section}}}
\newcommand{\Appendix}[1]{
\refstepcounter{section}
\begin{flushleft}
{\large\bf Appendix \thesection: #1}
\end{flushleft}}

\def\mpsl{\mathfrak{psl}}
\def\mD{\mathfrak D}
\def\mq{\mathfrak q}
\def\mpf{\mathfrak p} 
\def\Pr{{\mathbb P}}
\def\mU{\mathfrak U}
\def\msl{\mathfrak {sl}} 
 
\def\mg{\mathfrak g} 

\def\mH{\mathfrak H}
\def\mR{\mathfrak R}
\def\mL{\mathfrak L}
\def\mQ{\mathfrak Q}
\def\mS{\mathfrak S}
\def\msu{\mathfrak{su}}
\def\msl{\mathfrak{sl}}
\def\mE{\mathfrak E}
\def\mF{\mathfrak F}
\def\mH{\mathfrak H}
\def\mC{\mathfrak C}
\def\mP{\mathfrak P}
\def\mK{\mathfrak K}

\def\o{\theta}

\def\s{\sigma}

\def\C{\Gamma}

\newcommand{\Tr}{\operatorname{tr}}
\newcommand{\IM}{\operatorname{Im}}

\def\B0{{\boldsymbol 0}}

\def\C{{\mathbb C}}

\def\R{{\mathbb R}}

\newcommand{\Be}{\boldsymbol{e}}

\def\Dbarslash{\,\,{\raise.15ex\hbox{/}\mkern-12mu {\bar D}}}
\def\Dslash{\,\,{\raise.15ex\hbox{/}\mkern-12mu D}}
\def\delslash{\,\,{\raise.15ex\hbox{/}\mkern-9mu \partial}}
\def\delbarslash{\,\,{\raise.15ex\hbox{/}\mkern-9mu {\bar\partial}}}

\def\ket#1{\left| #1\right\rangle}

\def\bP{{\bf P}} 
\newcommand{\Sp}{{\color{white}()}}

\newcommand{\vpo}{\varphi_1}
\newcommand{\vpt}{\varphi_2}

\newcommand{\MAT}[1]{\begin{pmatrix} #1\end{pmatrix}}

\newcommand{\EQ}[1]{\begin{equation}\begin{split} #1
\end{split}\end{equation}}

\newcommand{\ALAT}[2]{\begin{alignat}{#1} #2
\end{alignat}}
\newcommand{\coset}[2]{\raisebox{.5ex}{$#1$}\Big/\raisebox{-.5ex}{$#2$}}
\newcommand{\FIG}[1]{\begin{figure}\begin{center} #1 \end{center}\end{figure}}

\title{A Relativistic Relative of the Magnon S-Matrix}
\author[a]{Ben Hoare,}
\author[b]{Timothy J. Hollowood}
\author[c]{and J. Luis Miramontes}
\affiliation[a]{Theoretical Physics Group, Blackett Laboratory, Imperial College, London SW7 2AZ, U.K.}
\affiliation[b]{Department of Physics, Swansea University, Swansea, SA2 8PP, U.K.}
\affiliation[c]{Departamento de F\'\i sica de Part\'\i culas and IGFAE,
Universidad
de Santiago de Compostela, 15782 Santiago de Compostela, Spain}

\emailAdd{benjamin.hoare08@imperial.ac.uk} 
\emailAdd{t.hollowood@swansea.ac.uk}
\emailAdd{jluis.miramontes@usc.es}

\abstract{We construct a relativistic scattering theory based on a $q$ deformation and large string tension limit of the magnon S-matrix of the string world sheet theory in $\text{AdS}_5\times S^5$. The S-matrix falls naturally into a previously studied class associated to affine quantum groups, in this case for a twisted affine loop superalgebra associated to an outer automorphism of 
$\msl(2|2)$. This infinite algebra includes the 
celebrated triply extended superalgebra $\mathfrak{psl}(2|2)\ltimes{\mathbb R}^3$, but only two of the centres, the lightcone components of the 2-momentum, are non-vanishing. The algebra has the interpretation as an extended supersymmetry algebra including a non-trivial R-symmetry. The representation theory of this algebra has some complications in that tensor products are reducible but indecomposable; however, we find that structure meshes perfectly with the bootstrap, or fusion, equations of S-matrix theory. The bootstrap equations can then be used inductively to generate the complete S-matrix. Unlike the magnon theory, the relativistic theory only has a finite set of states and we find that --- at least when the deformation parameter $q$ is a root of unity --- the spectrum matches 
precisely the soliton spectrum of the relativistic theory underlying the Pohlmeyer reduction of the string world sheet theory known as the semi-symmetric space sine-Gordon theory.}

\notoc
\begin{document}

\begin{flushright}   \small Imperial/TP/11/BH/02
\end{flushright}
\vspace{0.5cm} 

\maketitle

\newpage

\section{Introduction}\label{s1}

\pgfdeclarelayer{background layer} 
\pgfdeclarelayer{foreground layer} 
\pgfsetlayers{background layer,main,foreground layer}

There have been many remarkable implications of the fact that the Green-Schwarz superstring world-sheet theory defined on a $\text{AdS}_5\times S^5$ background is an integrable theory.\footnote{There is a large literature on this subject, see for example the series of review articles \cite{Beisert:2010jr} and references therein.}
Integrable field theories have some characteristic features: at the classical level, for instance, they exhibit the remarkable property of admitting more than one Hamiltonian formulation. For example, the mKdV integrable system, which is the simplest analogue of the string theory system, admits two distinct Poisson brackets, one of which is relativistically invariant. In fact, the latter is the formulation of the integrable system as the sine-Gordon theory. The two Poisson brackets are compatible, or  ``coordinated", so that one can actually write down an interpolating family of Poisson brackets. The string world sheet theory for  $\text{AdS}_5\times S^5$ mirrors this structure precisely. The string theory Poisson bracket structure is coordinated with the Poisson bracket of a relativistic system which appears as the Pohlmeyer reduction of the world-sheet theory \cite{Mikhailov:2005qv,Mikhailov:2005sy,Schmidtt:2010bi,Schmidtt:2011nr}. 
This Pohlmeyer reduced form of the $\text{AdS}_5 \times S^5$ superstring \cite{Grigoriev:2007bu} has received recent attention due to its classical equivalence to the original Green-Schwarz superstring
\cite{Pohlmeyer:1975nb,Hofman:2006xt}, while possessing a relativistic two-dimensional Lorentz symmetry, preserving the integrability of the superstring world sheet theory and also the UV-finiteness \cite{Roiban:2009vh}.
The reduction procedure is a fermionic generalisation \cite{Grigoriev:2007bu, Mikhailov:2007xr, Grigoriev:2008jq} of the original relation between the classical $O(3)$ sigma model and the sine-Gordon model \cite{Pohlmeyer:1975nb} (for a review of the reduction of more general bosonic models see \cite{Miramontes:2008wt} and references therein).
For the case of the superstring on  $\text{AdS}_5 \times S^5$,
the relativistic theory is in a class known as the 
semi-symmetric space sine-Gordon (SSSSG) theories that are related to a superspace generalization of a symmetric space 
known as a 
semi-symmetric space \cite{Serg,Zarembo:2010sg}. The particular semi-symmetric space in question is the coset
\EQ{
\coset{PSU(2,2|4)}{Sp(2,2)\times Sp(4)}\ .
\label{sss1}
}
The bosonic part of this coset is $\text{AdS}_5\times S^5$ itself. In more detail, the SSSSG theory is a gauged WZW model for the coset
\EQ{
\coset{Sp(2,2)\times Sp(4)}{SU(2)^{\times4}}\ ,
\label{doo}
}
coupled in a particular way to a set of fermions, deformed by the addition of a potential which breaks conformal invariance. 

As is seemingly ubiquitous, the integrability of the theory is controlled by a twisted affine loop algebra. Such algebras are defined by a Lie algebra, say $\mathfrak f$, and a finite order automorphism $\sigma$, $\sigma^N=1$. The algebra is defined over a set of generators
\EQ{
{\cal L}(\mathfrak f,\sigma)=\bigoplus_{n\in\mathbb Z}z^n\otimes\mathfrak f_{n\ \text{mod}\ N}\ ,
\label{hee}
}
where $\mathfrak f_n\subset\mathfrak f$ are the eigenspaces $\sigma(\mathfrak f_n)=e^{2\pi in/N}\mathfrak f_n$.  The algebra takes the form 
\EQ{
[z^m\otimes u,z^n\otimes v]=z^{m+n}\otimes[u,v]\ .
}
In the present case, $\mathfrak f=\mathfrak{psl}(2,2|4)$ and $\sigma$ is an automorphism of order 4 \cite{Grigoriev:2008jq,Hollowood:2011fq,Schmidtt:2010bi}. The potential term in the action has the effect of breaking the large affine symmetry to a smaller subalgebra which, remarkably,
contains
\EQ{
{\cal L}\big(\mathfrak p\big(\msl(2|2)\oplus\msl(2|2)\big),\sigma\big)\subset{\cal L}\big(\mathfrak{psl}(2,2|4),\sigma)\ .
\label{zxy}
}
Elements of the this algebra are associated to conserved charges whose Lorentz spin equals one half the grade.
The zero-graded bosonic subalgebra of this is the Lie algebra of the group $SU(2)^{\times4}$ in the denominator of the WZW coset \eqref{doo}. This is the group that is gauged in the WZW model but its global part remains as a symmetry of the spectrum.
The algebra \eqref{zxy} has single elements of grade $\pm2$ whose associated conserved charges are the lightcone components of the 2-momentum. These elements are centres of the algebra and this implies that the
affine loop superalgebra \eqref{zxy}, contains as a subalgebra the elements of grade between $-2$ and $2$, that is Lorentz spins $(0,\pm\frac12,\pm1)$, which generate the finite centrally-extended Lie algebra\footnote{Here, and in the following, we use the notation $\mathbb  R^2=\mathbb R\oplus\mathbb R$.}
\EQ{
\big(\mathfrak{psl}(2|2) \oplus \mathfrak{psl}(2|2))\ltimes \mathbb{R}^2 \ .
\label{symalg}
}
The central elements here, are components of the  2-momentum graded $\pm2$. This algebra is a non-trivial $\mathcal N = (8,8)$ supersymmetry algebra of the theory which acts in a (mildly) non-local way on the Lagrangian fields of the theory \cite{Schmidtt:2011nr,Hollowood:2011fq,Goykhman:2011mq}. This non-locality motivates the idea that the algebra may become $q$ deformed in the quantum theory with $q\to1$ in the classical limit.
Lorentz transformations can naturally be included by extending the algebra to include the derivation (the operator that grades the elements). The supersymmetry algebra includes the bosonic symmetry $SU(2)^{\times4}$, the global part of the gauge group of the gauged WZW model, which plays the r\^ole of a non-abelian R-symmetry.

Generalizing methods from the analysis of bosonic theories \cite{Hollowood:2011fm, Hollowood:2010dt}, the classical solitons of the theory were constructed and quantized in \cite{Hollowood:2011fq} and were found to transform in short (or atypical) representations of the symmetry algebra \eqref{symalg} of dimension $4a\times 4a$ with a mass spectrum of the form
\EQ{
m_a=\mu\sin\left(\frac{\pi a}{2k}\right)\ ,\qquad a=1,2,\ldots,k\ .
\label{mass2}
}
Here, $k$ is the level of the WZW model which is assumed to be a positive integer. 
Notice that the spectrum of soliton states is naturally truncated and also the states of lowest mass $a=1$ are identified with the perturbative states of the theory. This latter point deserves to be highlighted because it may seem surprising at first that perturbative excitations are actually solitons, a point described in detail in \cite{Hollowood:2010dt}.

The appearance of the doubly extended superalgebra $\mpsl(2|2)\ltimes\mathbb R^2$  is intriguing because the triply extended superalgebra $\mpsl(2|2)\ltimes\mathbb R^3$ 
plays a central r\^ole in the integrability and the magnon S-matrix on the string theory side. The additional central term can simply be understood as arising from a conventional central extension of the affine loop superalgebra ${\cal L}(\msl(2|2),\sigma)$ which vanishes in the SSSSG theory.
All this evidence suggests that the over-arching algebraic structure that organizes both the integrability of the string and the SSSSG theory is a quantum group, or $q$, deformation of a centrally extended loop algebra that we will denote $\msl(2|2)^{(\sigma)}$, based on an affinization of $\msl(2|2)$ 
with twisting by the outer automorphism $\sigma$ of order 4.\footnote{Note that $\sigma^2$ is an inner automorphism \cite{Serg3,Arutyunov:2009ga} and so in the notation of \cite{Kac} this algebra would be denoted $\msl(2|2)^{(2)}$. However, it differs by an inner automorphism from the twisting described in \cite{Gould1}.
Another issue is that for the quotient algebra $\mpsl(2|2)$ the automorphism lies inside a continuous group of $SL(2,\mathbb C)$ automorphisms that are connected to the identity. However, for $\msl(2|2)$ this is no longer the case meaning that it is genuinely different from the untwisted algebra $\msl(2|2)^{(1)}$ \cite{Serg2}. We refer to $\msl(2|2)^{(\sigma)}$ as a centrally extended loop algebra because, since $\msl(2|2)$ is not simple (it is reductive) it is not clear whether it fits into the usual class of Kac Moody superalgebras: for instance it has an infinite number of centres that span a Heisenberg subalgebra.}
This algebra includes the  triply extended superalgebra 
$\mpsl(2|2)\ltimes\mathbb R^3$ as a finite subalgebra.\footnote{More precisely we need two copies of this algebra as in \eqref{symalg} with central terms identified.} 
Roughly speaking, the string and SSSSG theories involve a quantization of the classical loop algebra ${\cal L}(\msl(2|2),\sigma)$ of two different kinds, the former by generating a non-trivial central charge and the latter by $q$ deformation.
This implies that there should be an interpolating non-relativistic theory, with both a central charge and $q$ deformation, from which the  
string theory is obtained by taking a limit $q\to1$ and the relativistic SSSSG theory by taking the affine central charge to 0. In fact, given that we identify the deformation parameter as
\EQ{
q=e^{i\pi/k}\ ,
\label{rrd}
}
where $k$ is the level of the WZW model,
the former is obtained in the limit $k\to\infty$, the classical limit of the SSSSG theory, and the latter in classical large tension limit of the string world-sheet theory.
The fact that the SSSSG theory involves a quantum group is expected because it has already been shown that S-matrix of the (bosonic) symmetric space sine-Gordon theory associated to  $\mathbb CP^{n+1}$ involves the affine (loop) quantum group $U_q(\msl(n)^{(1)})$ \cite{Hollowood:2010rv}.

The relativistic S-matrix theory that we construct turns out to be fit neatly into a previously known class of S-matrices
associated to affine quantum groups $U_q(\hat\mg)$  \cite{deVega:1990av,Bernard:1990ys,Hollowood:1992sy,Hollowood:1993fj,Delius:1995tc,Gandenberger:1997pk}. It is interesting that there are also existing examples involving affine superalgebras $\mathfrak{osp}(2|2)^{(1)}$ \cite{Bassi:1999ua}. The particles transform in representations of the affine quantum loop group and the S-matrix elements are proportional to the R-matrix for the affine quantum loop group. In the present case, the appropriate affine superalgebra is \eqref{zxy} above. 
In particular, the particles transform in representations of the finite supersymmetry algebra
\eqref{symalg}. 

The construction of the S-matrix begins with the 
R-matrix associated to the quantum group 
$U_q(\mathfrak{psl}(2|2) \ltimes \mathbb R^3)$ which 
was constructed in \cite{Beisert:2008tw}. This R-matrix is the $q$-deformation of the R-matrix that lies behind the magnon S-matrix. The first hint that this is the correct R-matrix is that a particular classical relativistic limit of this R-matrix identified in \cite{Beisert:2010kk} bears a remarkable resemblance to the tree-level S-matrix of the reduced $\text{AdS}_5 \times S^5$, or SSSSG, theory \cite{Hoare:2009fs}. In \cite{Hoare:2011fj} this relativistic limit was extended to all orders in the coupling and the resulting S-matrix proposed as a candidate for the S-matrix of the Lagrangian field excitations of the theory. In particular, the minimal CCD factor fixed by unitarity and crossing symmetry agrees with a perturbative computation. However, the origin of the quantum deformation in the perturbative computation is still an open question but not surprising given that the 
quantum groups play an important r\^ole in the quantization of WZW models \cite{Bernard:1990ys,Alekseev:1990vr,Caneschi:1996sr,Gawedzki:1990jc,Alekseev:1992wn}.

In \cite{Hoare:2011fj} the one-loop perturbative S-matrix for the reduced $\text{AdS}_3 \times S^3$ and the reduced $\text{AdS}_5 \times S^5$ theories were computed.\footnote{Similar computations for bosonic models were investigated in \cite{Hoare:2010fb}.} For the reduced $\text{AdS}_3 \times S^3$ theory it was found that (with the addition of a suitable local counterterm) the perturbative S-matrix satisfies the Yang-Baxter equation and is invariant under a $\mathcal N = (4,4)$ quantum-deformed supersymmetry. Due to the non-abelian nature of the $SU(2)^{\times4}$ gauge group for the reduced $\text{AdS}_5 \times S^5$  theory, the one-loop result did not satisfy the Yang-Baxter equation. However, motivated by the reduced $\text{AdS}_3 \times S^3$ example it was conjectured that the physical symmetry of the theory should be given by a quantum-deformation of \eqref{symalg}. A similar quantum-deformation is conjectured to occur in related bosonic models, examples of which are discussed in \cite{Hollowood:2010rv, Hollowood:2009sc, Hollowood:2009tw}.

This paper is organized as follows. In section \ref{s2}, we review some aspects of the S-matrix theories associated to affine quantum groups that will be useful in generalising to the superalgebra case of interest in the present paper.
In particular, we highlight some important issues concerning the bootstrap/fusion procedure, whereby certain simple poles of the S-matrix on the physical strip in rapidity space correspond to bound states in the direct or crossed channel. These bound states are then added to the spectrum and their S-matrix elements can then be deduced by the fusion equations. In section \ref{s3}, we draw heavily on 
\cite{Beisert:2008tw} and review the construction of the quantum group $U_q(\mathfrak{psl}(2|2) \ltimes \mathbb R^3)$ and discuss the relativistic limit identified in \cite{Beisert:2010kk,Hoare:2009fs}. In this section, the interpretation of the triply extended algebra as a subalgebra of a central extension of the affine loop superalgebra ${\cal L}(\msl(2|2),\sigma)$ is discussed and the magnon representation and relativistic soliton representations are then compared in some detail. 
In section \ref{s4}, we turn to the 
representation theory of the quantum-deformed superalgebra which is key to solving the bootstrap programme. In particular, the representation theory must mesh precisely with the analytic structure of the S-matrix in order to have a consistent S-matrix theory. 
Unfortunately the representation theory of the quantum-deformed algebra has not been investigated in detail. Experience with low dimensional representations suggests that, just as for a $q$ deformation of an ordinary Lie algebra, the representations are simple deformations of the representations of the undeformed algebra --- at least when $q$ is not a root of unity. Pending a more detailed investigation, we will assume that it is true and so we will review the representation theory of the undeformed superalgebra in some detail
based, in particular, on \cite{Zhang:2004qx}. A novel feature of the representation theory, is that tensor products are reducible but indecomposible and this feature will require careful treatment when we turn to the S-matrix. This we do in section 
\ref{s5} where we write down an S-matrix for the basic excitations transforming in the four-dimensional evaluation representation of $U_q(\mathfrak{sl}(2|2)^{(\sigma)})$ based on the R-matrix of \cite{Beisert:2008tw} appended with a suitable scalar, or CDD, factor to ensure unitarity and crossing. In section \ref{s6}, we turn to the question of bound states and the bootstrap programme. We show that the behaviour of tensor products of the quantum supergroup, including all the complications of indecomposable representations, meshes perfectly with the bootstrap/fusion procedure of S-matrix theory. 
The resulting bootstrap procedure gives a mass spectrum that precisely matches the semi-classical mass spectrum of the solitons of the SSSSG theory in \eqref{mass2} providing strong evidence that the tensor product of two copies of our relativistic S-matrix with an appropriate scalar factor is the S-matrix for the soliton excitations --- including the perturbative modes --- of the reduced theory.

We conclude the paper with a discussion of the closure of the bootstrap procedure and other open questions.  The closure is key to defining a consistent quantum S-matrix and 
we suggest a number of ways it which it could happen. The most 
cogent possibility is that the S-matrix theory requires that $k$, which is the level of the WZW model, is a positive integer and so the deformation parameter $q$ in \eqref{rrd} is a root of unity. Although, the representation theory of $U_q(\mathfrak{psl}(2|2)\ltimes\mathbb R^3)$  with $q$ a root of unity has not been investigated in any detail, experience with ordinary quantum groups suggests that the effect is to restrict the set of allowed representations and this would provide a mechanism for truncating the spectrum of states as indicated in \eqref{mass2}.

\section{Quantum Group S-Matrices}\label{s2}

S-matrix theories with symmetries that are associated to affine quantum groups arising as deformations of affine Lie algebras have been studied in the past \cite{deVega:1990av,Hollowood:1992sy,Hollowood:1993fj,Delius:1995tc,Gandenberger:1997pk}. The extension to the case of the affine superalgebra  $\mathfrak{osp}(2|2)^{(1)}$ appears in \cite{Bassi:1999ua}. In this section, we review, following loosely the approach described in \cite{Delius:1995tc}, some of the features of this body of work that will assist in applying the construction to our Lie superalgebra case.

The general setting involves the quantum group deformation $U_q(\hat\mg)$ of the universal enveloping algebra of an affine Lie algebra $\hat\mg$. The quantum group is defined by the Chevalley generators $\mH_j$, $\mE_j$, $\mF_j$, $j=0,1\ldots,r$, which have non-vanishing commutators
\EQ{
[\mH_j,\mE_k]=A_{jk}\mE_k\ ,\qquad[\mH_j,\mF_k]=A_{jk}\mF_k\ ,\qquad
[\mE_j,\mF_k]=\delta_{jk}[\mH_j]_{q_j}\ .
\label{ola}
}
They also obey quantum Serre relations that we will not write. In the above, $A_{jk}$ is the Cartan matrix of $\hat\mg$, and $q_j=q^{d_j}$, where $d_j$ are coprime intergers such that $d_jA_{jk}$ is symmetric. In the above,
\EQ{
[x]_q=\frac{q^x-q^{-x}}{q-q^{-1}}\ .
}

The QFT has particle multiplets of masses $m_a$ whose Hilbert spaces $V_a(\theta)$ are modules for certain finite dimensional unitary representations $\pi_a$ of $U_q(\hat\mg)$ with vanishing central charge.\footnote{Note that the representations in question are {\it not\/} unitary highest weight representations of the affine algebra, since such representations only exist when the centre of the affine algebra is a positive integer. Rather they are {\it evaluation representations\/} that lift from the finite Lie algebra $\mg$ to the loop algebra realization of $\hat\mg$.} The representations and associated modules are labelled by the rapidity $\theta$, which is associated algebraically to a gradation of $\hat\mg$ defined by a set of real numbers $\{s_j\}$. The representation with rapidity is then defined by 
\EQ{
\pi_a^\theta(\mE_j)=e^{s_j\theta}\pi_a(\mE_j)\ ,\qquad\pi_a^\theta(\mF_j)= e^{-s_j\theta}\pi_a(\mF_j)\ .
}
The quantum group has an associated co-product $\Delta$ which describes how the generators act on a tensor product:
\EQ{
\Delta(\mathfrak\mH_j)&=\mH_j\otimes1+1\otimes\mH_j\ ,\\
\Delta(\mathfrak E_j)&=\mathfrak E_j\otimes 1+q^{-\mathfrak H_j}\otimes \mathfrak E_j\ ,\\
\Delta(\mathfrak F_j)&=\mathfrak F_j\otimes q^{\mathfrak H_j}+1\otimes\mathfrak F_j\ .
\label{cop}
}
This can be used to define the representation on a multi-particle state; for example for two particles
\EQ{
\pi^{\theta_1\theta_2}_{ab}(u)=(\pi^{\theta_1}_a\otimes \pi^{\theta_2}_b)\,\Delta(u)\ ,\qquad u\in U_q(\hat\mg)\ .
}

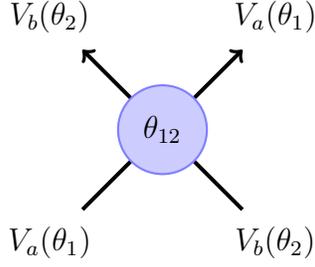
\begin{figure}
\begin{center}
\begin{tikzpicture} [line width=1.5pt,inner sep=2mm,
place/.style={circle,draw=blue!50,fill=blue!20,thick}]
\begin{pgfonlayer}{foreground layer}
\node at (1.5,1.5) [place] (sm) {$\theta_{12}$}; 
\end{pgfonlayer}
\node at (0,0) (i1) {$V_a(\theta_1)$};
\node at (3,0) (i2) {$V_b(\theta_2)$};
\node at (0,3) (i3) {$V_b(\theta_2)$};
\node at (3,3) (i4) {$V_a(\theta_1)$};
\draw[->] (i1) -- (i4);
\draw[->] (i2) -- (i3);
\end{tikzpicture}
\caption{\small The basic 2-body S-matrix elements that intertwine a tensor product of particle Hilbert spaces.}
\label{f1} 
\end{center}
\end{figure}

The S-matrix of a relativistic integrable theory are determined by the 2-body S-matrix elements $S_{ab}(\theta_{12})$, which act as intertwiners between the incoming and outgoing Hilbert spaces:
\EQ{
S_{ab}(\theta_{12}):\quad V_a(\theta_1)\otimes V_b(\theta_2)\longrightarrow V_b(\theta_2)\otimes V_a(\theta_1)\ ,
}
where $\theta_{12}=\theta_1-\theta_2$.
This illustrated in figure~\ref{f1}. 

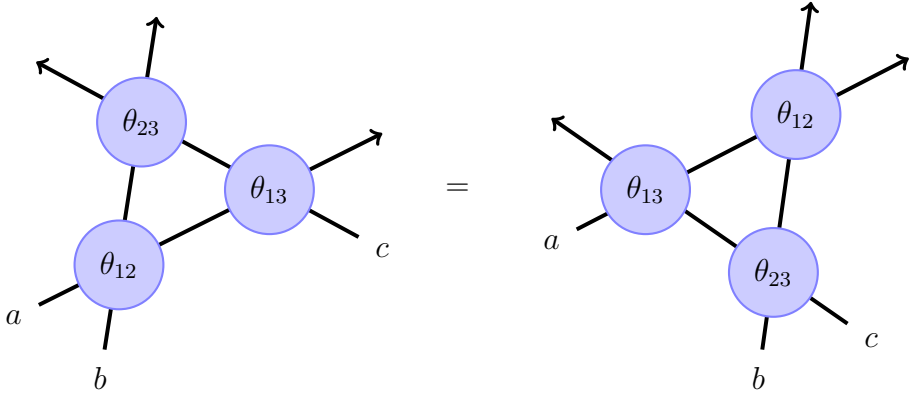
\begin{figure}
\begin{center}
\begin{tikzpicture} [line width=1.5pt,inner sep=2mm,
place/.style={circle,draw=blue!50,fill=blue!20,thick}]
\begin{pgfonlayer}{foreground layer}
\node at (1.5,1.5) [place] (sm1) {$\theta_{12}$}; 
\node at (1.8,3.4) [place] (sm2) {$\theta_{23}$}; 
\node at (3.5,2.5) [place] (sm3) {$\theta_{13}$}; 
\node at (8.5,2.5) [place] (sm4) {$\theta_{13}$}; 
\node at (10.2,1.4) [place] (sm5) {$\theta_{23}$}; 
\node at (10.5,3.5) [place] (sm6) {$\theta_{12}$}; 
\end{pgfonlayer}
\node at (6,2.5) {$=$};
\node at (10,0) (i1) {$b$};
\draw[->] (i1) -- (10.7,5);
\node at (11.5,0.5) (i2) {$c$};
\draw[->] (i2) -- (7.25,3.45);
\node at (7.25,1.8) (i3) {$a$};
\draw[->] (i3) -- (12,4.25) ;
\node at (5,1.7) (i4) {$c$};
\draw[->] (i4) -- (0.4,4.2);
\node at (0.1,0.8) (i5) {$a$};
\draw[->] (i5) -- (5,3.25);
\node at (1.25,0) (i6) {$b$};
\draw[->] (i6) -- (2,4.8);
\end{tikzpicture}
\caption{\small 
The Yang-Baxter Equation follows from the locality of interactions for widely separated wave packets and the fact that higher spin conserved charges can be used to shift the trajectories of the wavepackets without affecting the S-matrix.}
\label{f2}  
\end{center}
\end{figure}

For consistency the 2-body S-matrix element must satisfy the Yang-Baxter Equation
\EQ{
\big(S_{bc}(\theta_{23})\otimes1)&\big(1\otimes S_{ac}(\theta_{13})\big)\big(S_{ab}(\theta_{12})\otimes1)\\ &
=\big(1\otimes S_{ab}(\theta_{12})\big)\big(S_{ac}(\theta_{13})\otimes1)\big(1\otimes S_{bc}(\theta_{23})\big)\ ,
}
which is illustrated in figure~\ref{f2}.

The theory is invariant under the affine quantum symmetry in the sense that 
\EQ{
\pi^{\theta_2\theta_1}_{ba}(u)S_{ab}(\theta_{12})=
S_{ab}(\theta_{12})\pi^{\theta_1\theta_2}_{ab}(u)\ , \qquad u\in U_q(\hat\mg)\ .
}
So the S-matrix elements lie in the commutant of $U_q(\hat\mg)$ acting on a tensor product representation. It is often useful to write
\EQ{
S_{ab}(\theta)=X_{ab}(\theta)\check R_{ab}(x=e^{\lambda\theta})\ ,
}
where $X_{ab}(\theta)$ is a scalar factor which carries the important analytic structure of the S-matrix, in particular all the bound state poles, and $\check R_{ab}(x)$ is the quantum group $R$-matrix that carries all the tensorial structure.

The S-matrix of a relativistic QFT has to satisfy the two important conditions of unitarity and crossing symmetry. Unitarity requires
\EQ{
S_{ab}(\theta)S_{ba}(-\theta)=1\otimes1\ .
\label{unit}
}
Note that this is not the same as unitarity of the underlying QFT. This latter form of unitarity is intimately related to the behaviour of the S-matrix at bound state singularities as we explain below.
Crossing symmetry relies on the fact that each particle multiplet $V_a$ has a degnerate anti-particle multiplet $V_{\bar a}$, which transforms in a conjugate representation of $U_q(\hat\mg)$. For real representations, it is possible to have $\bar a=a$. Charge conjugation is then an invertible  map
\EQ{
{\cal C}:\quad V_a\longrightarrow V_{\bar a}
}
and then crossing symmetry requires
\EQ{
S_{ab}(\theta)=({\cal C}^{-1}\otimes1)\big(\sigma\cdot S_{\bar ba}(i\pi-\theta)\big)^{t_1}\cdot\sigma\cdot(1\otimes{\cal C})\ ,
\label{cs}
}
where $\sigma$ is the permutation on the tensor product $\sigma(V_a\otimes V_b)=V_b\otimes V_a$ and $t_1$ means transpose on first space in the tensor product which is well defined because $\sigma\cdot S_{ba}(\theta)\in\text{End}(V_b\otimes V_a)$. Charge conjugate relies in an algebraic sense on the antipode operation of the quantum group.

Let $\mg_0$ be the zero graded Lie subalgebra of  $\hat\mg$. In many cases, the modules $V_a$ are just finite dimensional irreducible representations of $\mg_0$, but there are situations in which a finite dimensional representation of $\mg_0$ cannot be lifted to $\hat\mg$
--- it is not {\it affinizable\/} --- as we shall highlight later. In these cases, $V_a$ is a reducible representation of $\mg_0$. 

The most non-trivial aspect of S-matrix theory is the analytic structure and its explanation in terms of bound states and anomalous thresholds. Bound states give 
rise to simple poles of the S-matrix on the physical strip, the region $0<\IM\,\theta<\pi$. For integrable field theories these poles occur at purely imaginary values:
if a bound state corresponding to 
particle $V_c$ is exchanged
in the direct channel then $S_{ab}(\theta)$, $\theta=\theta_{12}$, has a simple pole at the imaginary value $\theta=iu_{ab}^c$, with $0<u_{ab}^c<\pi$ where
\EQ{
m_c^2=m_a^2+m_b^2+2m_am_b\cos u_{ab}^c\ .
}
Note that if $c$ is a bound state of $a$ and $b$, then $a$ is a bound state of $c$ and $\bar b$, the anti-particle of $b$, and $b$ is a bound state of $c$ and $\bar a$,
\EQ{
u_{ab}^c+u_{b\bar c}^{\bar a}+u_{a\bar c}^{\bar b}=2\pi\ ,
}
as illustrated in figure~\ref{f3} 
\FIG{
\begin{tikzpicture} [line width=1.5pt,inner sep=2mm,
place/.style={circle,draw=blue!50,fill=blue!20,thick},proj/.style={circle,draw=red!50,fill=red!20,thick}]
\node at (2,2) [proj] (p1) {};
\node at (0.4,0.4) (i1) {$a$};
\node at (3.6,0.4) (i2)  {$b$};
\node at (2,4) (i3) {$c$};
\draw[->]  (i1) -- (p1);
\draw[->]  (i2) -- (p1);
\draw[<-]  (i3) -- (p1);
\node at (2,0.3) {$u_{ab}^c$};
\node at (0.8,2.5) {$u_{a\bar c}^{\bar b}$};
\node at (3.2,2.5) {$u_{b\bar c}^{\bar a}$};
\end{tikzpicture}
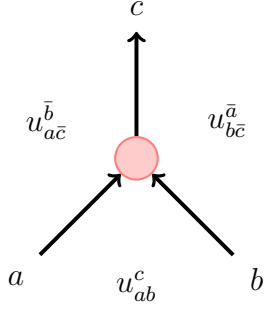
\caption{\small The rapidity angles for the 3-point functions.}
\label{f3}
}

The position of the bound-state poles must mesh with the representation theory of the quantum affine algebra. For generic values of the rapidities, the representation $\pi^{\theta_1\theta_2}_{ab}$ is irreducible. So, for consistency, at the bound-state pole $\theta_{12}=iu_{ab}^c$ the representation must become reducible and contain $V_c$ as a component. At this special point, if we write for $U_q(\mg_0)$ representations
\EQ{
V_a\otimes V_b=V_c\oplus V_c^\perp\ ,
}
then we require that $V^\perp_c$ lies in the kernel of $\text{Res}\,S_{ab}(iu_{ab}^c)$:
\EQ{
\text{Res}\,S_{ab}(iu_{ab}^c):\; V_c^\perp\longrightarrow0\ .
\label{ipm}
}
In general, the affinizable representation $V_c$ is reducible under $U_q(\mg_0)$. Suppose we write the decomposition as 
\EQ{
V_c=\oplus_jV_c^{(j)}\ ,
}
then near the pole we have
\EQ{
S_{ab}(\theta)\thicksim \frac i{\theta-iu_{ab}^c}\sum_j\rho_j\,\Pr_{ab}^{c,j}\ ,
\label{fgf}
}
where $\Pr_{ab}^{c,j}$ is the $U_q(\mg_0)$ invariant intertwiner which is only 
 non-vanishing on $V_c^{(j)}\subset V_a\otimes V_b$. It can be expressed as 
\EQ{
\Pr_{ab}^{c,j}=\EuScript P^{ba}_{c,j}\,\EuScript P_{ab}^{c,j}\ ,
}
where $\EuScript P_{ab}^{c,j}:\ V_a\otimes V_b\rightarrow V_c^{(j)}$ and $\EuScript P^{ba}_{c,j}:\ V^{(j)}_c\rightarrow V_b\otimes V_a$. We remark that when $a=b$, $\Pr_{aa}^{c,j}$ is a projection operator. 
In \eqref{fgf}, the numbers $\rho_j$ are required to be real, and unitarity of the underlying QFT dictates the sign. In simple cases, the sign is related to the parity of the bound state as found by Karowski~\cite{Karowski:1978ps}. For our S-matrix, the issue of unitarity and the signs of the residues is left for future analysis. The coupling of asymptotic states to the bound state is illustrated in figure~\ref{f4}. 
\FIG{
\begin{tikzpicture} [line width=1.5pt,inner sep=2mm,
place/.style={circle,draw=blue!50,fill=blue!20,thick},proj/.style={circle,draw=red!50,fill=red!20,thick}]
\begin{pgfonlayer}{foreground layer}
\node at (1,1) [proj] (p1) {};
\node at (1,3) [proj] (p2) {};
\node at (-0.5,-0.5) (i1) {$a$};
\node at (2.5,-0.5) (i2) {$b$};
\node at (-0.5,4.5) (i3) {$b$};
\node at (2.5,4.5) (i4) {$a$};
\node at (0.5,2) {$c$};
\node at (3,2.8) (s1) {$\sum_j\sqrt{|\rho_j|}\,\EuScript P^{ba}_{c,j}$};
\draw[->,thin] (s1) -- (p2);
\node at (3,1.2) (s3) {$\sum_j\sqrt{|\rho_j|}\,\EuScript P_{ab}^{c,j}$};
\draw[->,thin] (s3) -- (p1);
\end{pgfonlayer}
\draw[-] (i1) -- (p1);
\draw[-] (i2) -- (p1);
\draw[<-] (i3) -- (p2);
\draw[<-] (i4) -- (p2);
\draw[-,red] (p2) -- (p1);
\end{tikzpicture}
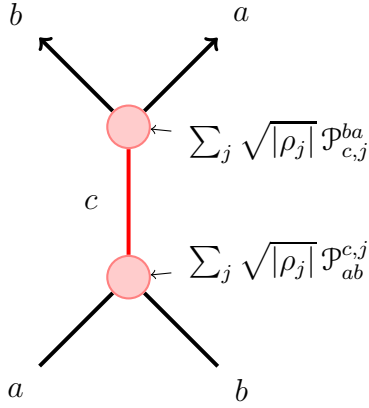
\caption{\small The anatomy of the S-matrix in the vicinity of a bound state pole. The coupling of the bound state to the asymptotic states involves the maps $\EuScript P_{ab}^{c,j}$ and $\EuScript P_{c,j}^{ba}$ as well as the weightings $|\rho_j|$. }
\label{f4}
}

The fact that $c$ can appear as a bound state of $a$ and $b$ means that the S-matrix elements of $c$ with other states, say $d$, can be written in terms of those of $a$ and $b$. This is the essence of the bootstrap, or fusion, programme. The relation between the S-matrix elements can be written concretely as
\EQ{
S_{dc}(\theta)&=\left(\sum_j\sqrt{|\rho_j|}\,\EuScript P_{ab}^{c,j}\otimes1\right) \Big(1\otimes S_{db}(\theta+i\bar u_{b\bar c}^{\bar a})\Big) \\ &\qquad\times\left(S_{da}(\theta-i\bar u_{a\bar c}^{\bar b})\otimes1\right)  \left(1\otimes\sum_l\frac1{\sqrt{|\rho_l|}}\,\EuScript P_{c,l}^{ab}\right)\ ,
\label{bfs}
}
where $\bar u_{ab}^c=\pi-u_{ab}^c$. This expression follows in an obvious way from the equality illustrated in figure \ref{f5}.  
\FIG{
\begin{tikzpicture} [scale=0.7,line width=1.5pt,inner sep=2mm,
place/.style={circle,draw=blue!50,fill=blue!20,thick},proj/.style={circle,draw=red!50,fill=red!20,thick}]
\begin{pgfonlayer}{foreground layer}
\node at (1.5,1.5) [place] (sm1) {}; 
\node at (1.8,3.4) [proj] (sm2) {}; 
\node at (3.5,2.5) [place] (sm3) {}; 
\node at (10.7,3.4) [place] (sm10) {}; 
\node at (11.5,1.7) [proj] (sm11) {};
\node at (6.5,2.5) {$=$};
\end{pgfonlayer}
\node at (11,0) (k1) {$a$};
\draw[-] (k1) -- (sm11);
\node at (13,0.7) (k2) {$b$};
\draw[-]  (k2) -- (sm11);
\node at (8.7,2.3) (k8) {$d$};
\draw[->] (k8) -- (12.8,4.5);
\node at (9.8,5.5) (k3) {$c$};
\draw[->,red] (sm11) -- (k3);
\node at (0.8,5.5) (k4) {$c$};
\draw[->,red] (sm2) -- (k4);
\node at (5,1.7) (k5) {$b$};
\draw[-] (k5) -- (sm2);
\node at (1.25,0) (k6) {$a$};
\node at (0,0.8) (k7) {$d$};
\draw[->] (k7) --  (4.7,3.15);
\draw[-] (k6) -- (sm2);
\node at (-1,2) (t1) {$\theta-i\bar u_{a\bar c}^{\bar b}$};
\draw[->,thin] (t1) -- (sm1);
\node at (3.5,4.3) (t2) {$\theta+i\bar u_{b\bar c}^{\bar a}$};
\draw[->,thin] (t2) -- (sm3);
\node at (10.2,2) (t3) {$\theta$};
\draw[->,thin] (t3) -- (sm10);
\node at (-1,3.5) (q1) {$\sum_j\sqrt{|\rho_j|}\,\EuScript P_{ab}^{c,j}$};
\node at (14.4,2.5) (q2) {$\sum_j\sqrt{|\rho_j|}\,\EuScript P_{ab}^{c,j}$};
\draw[->,thin] (q1) -- (sm2);
\draw[->,thin] (q2) -- (sm11);
\end{tikzpicture}
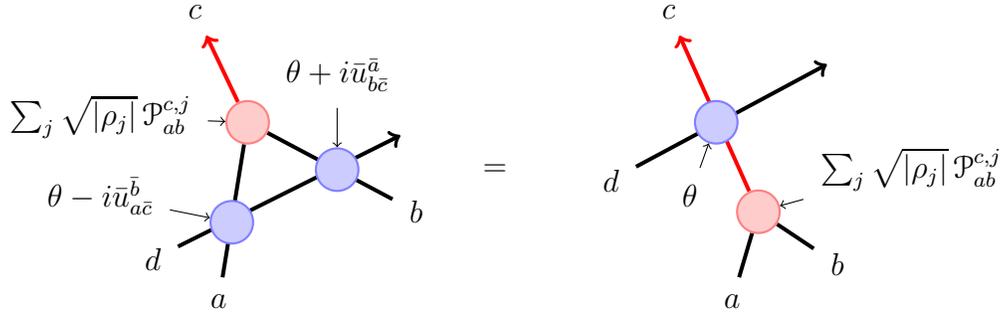
\caption{The bootstrap/fusion equations result from the equality of the diagrams above. One understands these diagrams in terms of localized wavepackets. The higher spin conserved charges implied by integrability can be used to move the trajectory of particle $d$ so that it either interacts with bound state $c$ or the particles $a$ and $b$ of which $c$ is composed. In order to isolate $S_{dc}(\theta)$ one has to act on the right by $1\otimes\sum_j\EuScript P^{ab}_{c,j}/\sqrt{|\rho_j|}$.}
\label{f5}
}
The most difficult aspect of building a consistent QFT is finding closure of the bootstrap programme; that is being able to account for all the poles of the S-matrix on the physical strip either in terms of direct- or cross-channel bound states, which lead to simple poles, or anomalous thresholds, which in $1+1$-dimensions manifest as poles of arbitrary order. 

\subsection{Examples}

The first example involves the affine algebra $\msl(n)^{(1)}$ with the homogeneous gradation where only $s_0\neq0$ and so $\mg_0=\msl(n)$. We will denote $x=e^{s_0\theta/2}$. The $R$-matrix for the tensor product $V_1\otimes V_1$, where $V_1$ is the vector representation of $\msl(n)$, can be written \cite{Jimbo}
\EQ{
\check R(x)=(xq-x^{-1}q^{-1})\Pr_++(x^{-1}q-xq^{-1})\Pr_-\ ,
}
where $\Pr_\pm$ are $U_q(\msl(n))$ invariant projectors onto the rank-2 symmetric and anti-symmetric representations. These projectors are related to a $q$ deformation of the symmetric group known as the Hecke algebra. The form of $\check R(x)$ above shows that there are special points when $x^2=q^{\pm2}$ where $\check R(x)$ becomes a projector onto the symmetric and anti-symmetric representations. Both these representations are affinizable. QFTs can be constructed which include either the symmetric or anti-symmetric representations in the spectrum by appropriate choices of $s_0$ and the scalar factor.

We now turn to an affine algebra $\mathfrak{so}(2n)^{(1)}$, again with the homogeneous gradation. In this case the $\check R(x)$ matrix for a tensor product $V_1\otimes V_1$, where $V_1$ is the vector representation takes the form \cite{Jimbo}
\EQ{
\check R(x)&=(xq-x^{-1}q^{-1})(xq^{n/4}-x^{-1}q^{-n/4})\Pr_+\\ &\qquad+(x^{-1}q-xq^{-1})(xq^{n/4}-x^{-1}q^{-n/4})\Pr_-\\ &\qquad\qquad+(x^{-1}q-xq^{-1})(x^{-1}q^{n/4}-xq^{-n/4})\Pr_\bullet\ ,
}
where $\bullet$ is the singlet. In this case, at the special points $x=\pm q^{-1}$ we have 
\EQ{
\check R(\pm q^{-1})&=(q^2-q^{-2})(q^{n/4-1}-q^{-n/4+1})\Pr_-\\ &\qquad+(q^2-q^{-2})(q^{n/4+1}-q^{-n/4-1})\Pr_\bullet\ ,
}
which is not a projector onto a single irreducible representation of $U_q(\mathfrak{so}(2n))$; rather the representation is reducible and the
$\check R$-matrix is a weighted sum of projection operators. This is symptom of the fact that anti-symmetric representation by itself is not ``affinizable", and in order to find an irreducible representation of the affine algebra one must take the reducible module $V_2=V_-\oplus V_\bullet$.
In contrast, notice that when $x=\pm q$ the $\check R$-matrix becomes a projector onto the symmetric representation.

\section{Centrally Extended $\mathfrak{psl}(2|2)$ and its Quantum Group}\label{s3}

In this section we describe the theory of the Lie superalgebra $\mathfrak{psl}(2|2)$, its central extensions and its $q$, or quantum group, deformation. We pay particular attention to the defining representation and the differences between the magnon and soliton representations.
In the following section, we turn to its rich representation theory. Since this algebra has been extensively studied, our discussion will not be comprehensive and, in particular, we shall draw extensively on the discussion by Beisert and Koroteev \cite{Beisert:2008tw} and use their notation throughout. 

The Lie superalgebra $\mathfrak{su}(2|2)$ is generated by $4\times 4$ anti-Hermitian matrices which, in $2\times2$ block notation, are of the form
\EQ{
M=-M^\dagger=\left(\begin{array}{c|c}m & \theta \\\hline \eta & n\end{array}\right)\ .
\label{vxx}
}
The matrices $m$ and $n$ are Grassmann even, with $m^\dagger=-m$ and $n^\dagger=-n$, and $\theta$ and $\eta$ are Grassmann odd, with $\eta=-\theta^\dagger$.\footnote{Here, $\dagger$ is the usual hermitian conjugation, $M^\dagger=(M^*)^t$, but with the definition that complex conjugation is anti-linear on products of Grassmann odd elements
\EQ{
(\theta_1\theta_2)^*=\theta_2^*\theta_1^*\ ,
}
which guarantees that $(M_1M_2)^\dagger=M_2^\dagger M_1^\dagger$.}
These matrices are required to have vanishing supertrace $\text{str}\,M=-\Tr\,m+\Tr\,n=0$. In most of the following, we will work with the complex form of the algebra $\msl(2|2)$.
Notice that the algebra includes the element ${\bf 1}$ as a centre, and this allows one to define  $\mathfrak{psl}(2|2)$ as the quotient $\mathfrak{sl}(2|2)/{\bf1}$. 

The Lie superalgebra $\mathfrak{psl}(2|2)$ is unique in that it can be extended with 3 independent centres. 
Using the notation of \cite{Beisert:2008tw}, the centrally extended algebra can be written as follows. 
The generators of $\mathfrak{psl}(2|2)$ consist of the $\mg_0=\mathfrak{sl}(2)\oplus\msl(2)$ even generators $\mathfrak R^a{}_b$ and $\mathfrak L^\alpha{}_\beta$, with traceless conditions $\mathfrak R^1{}_1=-\mathfrak R^2{}_2$ and $\mathfrak L^1{}_1=-\mathfrak L^2{}_2$, and the odd generators $\mathfrak Q^\alpha{}_b$ and $\mathfrak S^a{}_\beta$, the latter to be multiplied by Grassmann numbers to give an algebra element. The brackets of the algebra involving elements of $\mg_0$ are
\begin{alignat}{2}
\notag [\mR^a{}_b,\mR^c{}_d]&=\delta^c_b\mR^a{}_d-\delta^a_d\mR^c{}_b\ ,\qquad&
[\mL^\alpha{}_\beta,\mL^\gamma{}_\delta]&=\delta^\gamma_\beta
\mL^\alpha{}_\delta-\delta^\alpha_\delta
\mL^\gamma{}_\beta
\ ,\\ 
[\mR^a{}_b,\mQ^\gamma{}_d]&=-\delta^a_d\mQ^\gamma{}_b+\tfrac12
\delta^a_b\mQ^\gamma{}_d\ ,\qquad&
[\mL^\alpha{}_\beta,\mQ^\gamma{}_d]&=\delta^\gamma_\beta\mQ^\alpha{}_d-\tfrac12\delta^\alpha_\beta\mQ^\gamma{}_d\ ,\label{com1}\\
\notag[\mR^a{}_b,\mS^c{}_\delta]&=\delta^c_b\mS^a{}_\delta-\tfrac12\delta^a_b\mS^c{}_\delta\ ,&
[\mL^\alpha{}_\beta,\mS^c{}_\delta]&=-\delta^\alpha_\delta\mS^c{}_\beta+\tfrac12\delta^\alpha_\beta\mS^c{}_\delta\ ,
\end{alignat}
while the odd generators satisfy the anti-commutation algebra
\EQ{
\{\mQ^\alpha{}_b,\mS^c{}_\delta\}&=\delta^c_b\mL^\alpha{}_\delta+\delta^\alpha_\delta\mR^c{}_b+\delta^c_b\delta^\alpha_\delta\mC\ ,\\
\{\mQ^\alpha{}_b,\mQ^\gamma{}_d\}&=\varepsilon^{\alpha\gamma}\varepsilon_{bd}\mP\ ,
\qquad\{\mS^a{}_\beta,\mS^c{}_\delta\}=\varepsilon^{ac}\varepsilon_{\beta\delta}\mK\ .\label{com2}
}
In the above, $\mC$, $\mP$ and $\mK$ are the 3 central extensions which commute with all other generators.

The question before us 
is what is the relation to the defining 4-dimensional representation of the real form $\msu(2|2)$ described above. Introducing the basis $\Be_{ij}$, $i,j\in\{1,2,3,4\}$, as the matrix with a 1 in position $(i,j)$ and 0 elsewhere, the  $\mg_0$ generators  are simply
\begin{alignat}{3}
\notag \mR^1{}_1&=-\mR^2{}_2=
\tfrac12\big(\Be_{11}-\Be_{22}\big)\ ,\quad&\mR^1{}_2&=\Be_{12}\ ,\quad&\mR^2{}_1&=\Be_{21}\ ,\\
\mL^1{}_1&=-\mL^2{}_2=
\tfrac12\big(\Be_{33}-\Be_{44}\big)\ ,\quad&\mL^1{}_2&=\Be_{34}\ ,\quad&\mL^2{}_1&=\Be_{43}\ .
\end{alignat}
For the odd generators there is some freedom:
\begin{alignat}{2}
\notag\mQ^1{}_1&=a \Be_{31}+b\Be_{24}\ ,\qquad&\mQ^1{}_2&=a\Be_{32}-b\Be_{14}\ ,\\
\mQ^2{}_1&=a\Be_{41}-b\Be_{23}\ ,\qquad&\mQ^2{}_2&=a\Be_{42}+b\Be_{13}
\label{genA}
\end{alignat}
and 
\begin{alignat}{2}
\notag\mS^1{}_1&=d\Be_{13}+c\Be_{42}\ ,\qquad&\mS^1{}_2&=d\Be_{14}-c\Be_{32}\ ,\\
\mS^2{}_1&=d\Be_{23}-c\Be_{41}\ ,\qquad&\mS^2{}_2&=d\Be_{24}+c\Be_{31}\ ,
\label{genB}
\end{alignat}
where the parameters satisfy
\EQ{
ad-bc=1\ .
\label{con}
}
In this 4-dimensional representation, the three centres are all proportional to the identity matrix which is the centre of $\msl(2|2)$: $\mathfrak C=C\cdot{\bf 1}$, $\mP=P\cdot{\bf1}$ and $\mK=K\cdot{\bf1}$, where 
\EQ{
C=\tfrac12(ad+bc)\ ,\qquad P=ab\ ,\qquad K=cd
}
and due to \eqref{con} they are subject to the constraint
\EQ{
C^2-PK=\tfrac14\ .
\label{conL2}
}
Different choices for $\{a,b,c,d\}$ give rise to different representations of the algebra, for example later we will focus on two particular representations associated to the magnons of string theory and the solitons of the SSSSG theory. 

The $\mg=\mpsl(2|2)\ltimes\mathbb R^3$ algebra admits a ${\mathbb Z}$-gradation 
\ALAT{3}{
\notag&s(\mR^a{}_b)=s(\mL^\alpha{}_\beta)=0\ ,\qquad& s(\mQ^\alpha{}_b)&=1\ ,\qquad& s(\mS^a{}_\beta)&=-1\ ,\\
&s(\mC)=0\ ,\qquad& s(\mP)&=2\ ,\qquad& s(\mK)&=-2\ ,
\label{zgr}
}
which can be associated to 
an additional element that can be added to the algebra known as the  {\it derivation\/} $\mD$ which, for the basis elements $u\in\{\mR^a{}_b,\mL^\alpha{}_\beta,\mQ^\alpha{}_b,\mS^a{}_\beta,\mP,\mK,\mC\}$ of the algebra, acts as
\EQ{
[\mD,u]=s(u)u\ .
\label{soo}
}
In the relativistic soliton theory, the grade of an element will be identified with minus twice the Lorentz spin. In this interpretation, $\mQ^\alpha{}_b$ and $\mS^a{}_\beta$ have spins $\mp\tfrac12$ and so are interpreted as supersymmetry generators and $\mathfrak P$ and $\mathfrak K$ will be identified, up to an overall constant, with the lightcone components of the 2-dimensional momentum, of spin $\mp1$. Notice that, in comparison with the case in section \ref{s2}, the momentum generators are part of the symmetry algebra as one expects in a supersymmetric theory. The derivation is then the generator of Lorentz transformations.

It should be apparent that the algebra $\mg=\mpsl(2|2)\ltimes\mathbb R^3$ 
has the whiff of an affine algebra about it even though it is finitely generated. 
The finite set of elements have grades restricted to the interval $[-2,+2]$. 
The algebra $\mg$ can be thought of as a finite-dimensional subalgebra 
of the centrally extended loop superalgebra $\msl(2|2)^{(\sigma)}$
defined below, associated to a $\mathbb Z_4$ automorphism $\sigma$.
We do not have a complete understanding of the r\^ole of such an infinite algebra, but we can make the following observations.
The appearance of a $\mathbb Z_4$ automorphism is not a surprise since the semi-symmetric space \eqref{sss1} is defined by such an automorphism of the superalgebra $\mpsl(2,2|4)$. If we take the generators of $\msl(2|2)$ as $\mL^\alpha{}_\beta$, $\mR^a{}_b$, $\mQ^\alpha{}_b$, $\mS^a{}_\beta$ along with the unit matrix $\bf1$, then
the action of the automorphism $\sigma$ in this basis is simply related to the $\mathbb Z$ grade \eqref{zgr} by
\EQ{
\sigma(u)=e^{i\pi s(u)/2}u\ ,
\label{dwe}
}
along with $\sigma({\bf 1})=-{\bf1}$. Denoting the eigenspaces of the
algebra under $\sigma$, $\sigma(\mg_j)=e^{i\pi j/2}\mg_j$, we have
\ALAT{2}{
\notag \msl(2|2)_0&=\{\mR^a{}_b,\mL^\alpha{}_\beta\}\ ,\qquad&
\msl(2|2)_1&=\{\mQ^\alpha{}_b\}\ ,\\
\msl(2|2)_2&=\{{\bf1}\}\ ,\qquad&
\msl(2|2)_3&=\{\mS^a{}_\beta\}\,.
\label{identify}
}
The algebra $\msl(2|2)^{(\sigma)}$ is obtained as a central extension of the loop algebra ${\cal L}(\msl(2|2),\sigma)$, completed with a derivation:
\EQ{
\msl(2|2)^{(\sigma)}={\cal L}(\msl(2|2),\sigma)\oplus\mathbb C\,\mC\oplus\mathbb C\,\mD\ ,
}
which takes the form
\EQ{
&[z^m\otimes u,z^n\otimes v]=z^{m+n}\otimes[u,v]+m\,\text{str}(uv)\delta_{m+n,0}\,\mC\ ,\\
&[\mC,z^m\otimes u]=0\ ,\qquad[\mD,z^m\otimes u]=mz^m\otimes u\ ,\qquad[\mC,\mD]=0\ .
}
Notice that $\mC$, the third central term of $\mg$, is identified with the conventional central charge of the infinite algebra. Although it would be interesting to further investigate the structure of this infinite algebra, for our purposes it will be enough to deal with the much more manageable finite algebra 
$\mg=\mathfrak{psl}(2|2)\ltimes\mathbb R^3$ which is a finite subalgebra of $\msl(2|2)^{(\sigma)}$ consisting of all the elements of grades $[-2,+2]$ with $\mD$ identified with the derivation in \eqref{soo}. In particular, the unique elements of grade $\pm2$ are identified with the two centres $\mK$ and $\mP$, respectively:
\EQ{
z^2\otimes{\bf1}=\mP\ ,\qquad z^{-2}\otimes{\bf1}=\mK\ .
}
The fact that these elements are the only elements of grade $\pm2$ and they are in the centre of the algebra is the reason why $\mg$ is a closed finite subalgebra of the full infinite dimensional affine algebra. Before proceeding, we point out a connection with the affine algebra $\msl(2|2)^{(2)}$ discussed in \cite{Gould1}. The outer automorphism used to define this twisted affinization differs from ours by an inner automorphism. Consequently the affine algebras $\msl(2|2)^{(\sigma)}$ and $\msl(2|2)^{(2)}$ are isomorphic. However, the difference by an inner automorphism means that the algebras have different $\mathbb Z$ gradations. The difference in gradations has physical consequences, for instance, the zero graded algebra is $\msl(2)\oplus\msl(2)$ in our case, but $\mathfrak{osp}(2|2)\simeq\msl(2|1)$ for the gradation implicit in \cite{Gould1}.

The centrally extended Lie superalgebra $\mathfrak{psl}(2|2)\ltimes\mathbb R^3$ admits a group of outer automorphisms $SL(2, \mathbb C)$~\cite{Beisert:2006qh,Arutyunov:2009ga}  which acts on the Grassman odd generators $\mathfrak Q^\alpha{}_b$ and $\mathfrak S^a{}_\beta$ and, thus, on the central elements, leaving the combination
\EQ{
\vec{\mC}^2=\mC^2 - \mP\mK
}
invariant. It was used in~\cite{Beisert:2006qh} to construct representations of $\mathfrak{psl}(2|2)\ltimes \mathbb R^3$ for generic values of the three central elements in terms of the representations of $\mathfrak{sl}(2|2)$. In eqs.~\eqref{genA} and~\eqref{genB}, these automorphisms relate different choices of the parameters $\{a,b,c,d\}$ which lead to inequivalent realizations of the basis of generators. In particular, the action of the outer automorphism does not act in a way that is consistent with the $\mathbb Z$ grades of the generators \eqref{zgr}.

Before we describe the quantum group $U_q(\mg)$, it is helpful to introduce a Chevalley basis for the complex algebra  consisting of generators $\{\mathfrak E_i,\mathfrak F_i,\mathfrak H_i\}$. Following \cite{Beisert:2008tw}, we choose 
\begin{alignat}{3}
\notag\mE_1&=\mR^2{}_1\ ,\qquad&\mE_2&=\mQ^2{}_2\ ,\qquad&\mE_3&=\mL^1{}_2\ ,\\
\mF_1&=\mR^1{}_2\ ,\qquad&\mF_2&=\mS^2{}_2\ ,\qquad&\mF_3&=\mL^2{}_1\ ,
\label{pqo}
\end{alignat}
in which case
\EQ{
\mH_1=-2\mR^1{}_1\ ,\qquad\mH_2=-\mC-\tfrac12\mH_1-\tfrac12\mH_3\ ,\qquad\mH_3=-2\mL^1{}_1\ .
\label{wsw}
}
The Chevalley generators satisfy the algebra 
\EQ{
[\mH_i,\mE_j]=A_{ij}\mE_j\ ,\qquad[\mH_i,\mF_j]=-A_{ij}\mF_j
\label{bvv}
}
and 
\EQ{
[\mE_1,\mF_1]=\mH_1\ ,\qquad\{\mE_2,\mF_2\}=-\mH_2\ ,\qquad[\mE_3,\mF_3]=-\mH_3\ .
\label{buu}
}
In the above,
\EQ{
 A_{ij}=\left(\begin{array}{rrr}2 & \ -1 & 0\\ -1 & 0 & 1\\ 0 &  1 & \ -2\end{array}\right)
}
is the (degenerate) Cartan matrix of $\mg$. The remaining (anti-)commutators are written down in 
\cite{Beisert:2008tw}.

\subsection{The quantum deformation}\label{s3.1}

In order to proceed, we need to describe the 
quantum group deformation $U_q(\mg)$.
For the Chevalley generators, it corresponds to the deformation of  \eqref{buu} to
\EQ{
[\mE_1,\mF_1]=[\mH_1]_q\ ,\qquad\{\mE_2,\mF_2\}=-[\mH_2]_q\ ,\qquad[\mE_3,\mF_3]=-[\mH_3]_q\ .
\label{buu2}
}
In contrast, the commutators \eqref{bvv} are not modified, but it is convenient to write them in the exponentiated form
\EQ{
q^{\mH_i}\mE_j=q^{A_{ij}}\mE_jq^{\mH_i}\ ,\qquad
q^{\mH_i}\mF_j=q^{-A_{ij}}\mF_jq^{\mH_i}\ .
}
The Serre relations are also deformed although we shall not need the explicit expressions here. 
In the following, we take the deformation parameter\,\footnote{In \cite{Hoare:2011fj} the quantum deformation parameter was taken to be related to $k$ as $q=e^{-i\pi/k}$. This amounts to choosing the bosonic states of the factorized S-matrix to originate from the $\text{AdS}_5$ sector and the bound states to transform in the short representations $\langle 0, a \rangle$ (see section \ref{s4}). 
Here, to mirror the construction of the bound states in the superstring theory, we take the bosonic states of the factorized S-matrix to originate from the $S^5$ sector. Then, the bound states transform in the short representations $\langle a, 0 \rangle$, and correspondingly $q=e^{i\pi/k}$.\label{footnote1}} 
\EQ{
q=e^{i\pi/k}\ ,
}
where $k$ is a positive real number which we assume is not an integer. The case with $k$ an integer is considered briefly in section~\ref{s6.1}. 

At the level of the 4-dimensional representation we can achieve the $q$ deformation by modifying the Chevalley generators with appropriate factors of $q$:
\begin{alignat}{3}
\notag\mE_1&=q^{1/2}\mR^2{}_1\ ,\qquad&\mE_2&=\mQ^2{}_2\ ,\qquad&\mE_3&=q^{-1/2}\mL^1{}_2\ ,\\
\mF_1&=q^{-1/2}\mR^1{}_2\ ,\qquad&\mF_2&=\mS^2{}_2\ ,\qquad&\mF_3&=q^{1/2}\mL^2{}_1\ .
\label{onn}
\end{alignat}
As before, in the 4-dimensional representation, the centres are $P=ab$ and $K=cd$, but now the other centre is defined implicitly by
\EQ{
ad=[C+\tfrac12]_q\ ,\qquad bc=[C-\tfrac12]_q\ .
\label{ll1}
}
Then, the 
constraint  \eqref{con} is modified to 
\EQ{
(ad-qbc)(ad-q^{-1}bc)=1\ ,
\label{ll2}
}
which is equivalent to 
\EQ{
[C]_q^2-PK=[\tfrac12]_q^2\ .
\label{con2}
}

For later use we can write the action of the Chevalley generators on the 4-dimensional representation by introducing a basis $\{\ket{\phi^1},\ket{\phi^2},\ket{\psi^1},\ket{\psi^2}\}$ for the action of the basis matrices $\Be_{ij}$:
\ALAT{3}{
\notag &\mathfrak H_1\ket{\phi^1}=-\ket{\phi^1},\qquad &&\mathfrak E_1
\ket{\phi^1}=q^{1/2}\ket{\phi^2},\qquad&&\mathfrak F_2\ket{\phi^1}=c\ket{\psi^1},\\ 
\notag&\mathfrak H_1\ket{\phi^2}=\ket{\phi^2},\  &&\mathfrak E_2
\ket{\phi^2}=a\ket{\psi^2},\ &&\mathfrak F_1\ket{\phi^2}=q^{-1/2}\ket{\phi^1},\\ 
&\mathfrak H_3\ket{\psi^2}=\ket{\psi^2},\  &&\mathfrak E_3
\ket{\psi^2}=q^{-1/2}\ket{\psi^1},\ &&\mathfrak F_2\ket{\psi^2}=d\ket{\phi^2},\\ 
\notag&\mathfrak H_3\ket{\psi^1}=-\ket{\psi^1},\  &&\mathfrak E_2
\ket{\psi^1}=b\ket{\phi^1},\ &&\mathfrak F_3\ket{\psi^1}=q^{1/2}\ket{\psi^2}\ ,
\label{fppq}
}
where the action of $\mH_2$ is not written since it is determined by \eqref{wsw}. 
The Chevalley generators $\mE_j$ and $\mF_j$ have $\mathbb Z$ grades $\pm s_j$, respectively, with $s_j=(0,1,0)$.

A useful parameterization of $\{a,b,c,d\}$ for the quantum group was introduced in \cite{Beisert:2008tw}:
\ALAT{2}{
\notag a&=\sqrt g\gamma\ ,\qquad&
b&=\frac{\sqrt g\alpha}{\gamma}\left(1-q^{2C-1}\frac{x^+}{x^-}\right)\ ,\\
c&=\frac{i\sqrt g\gamma}\alpha\frac{q^{-C+1/2}}{x^+}\ ,\qquad &
d&=\frac{i\sqrt g}\gamma q^{C+1/2}\left(x^--q^{-2C-1}x^+\right)\ ,
}
subject to a constraint
\EQ{
\frac{x^+}q+\frac q{x^+}-qx^--\frac1{qx^-}+ig(q-q^{-1})\left(\frac{x^+}{qx^-}-\frac{qx^-}{x^+}\right)=\frac ig\ ,
\label{xll}
}
which follows from \eqref{ll1} and \eqref{ll2}.
The three centres are given by
\EQ{
q^{2C}&=q\frac{(q-q^{-1})/x^+-ig^{-1}}{(q-q^{-1})/x^--ig^{-1}}
=q^{-1}\frac{(q-q^{-1})x^++ig^{-1}}{(q-q^{-1})x^- + ig^{-1}}\ ,\\
P&=g\alpha\left(1-q^{2C}\frac{x^+}{qx^-}\right)\ ,\qquad
K=g\alpha^{-1}\left(q^{-2C}-\frac{qx^-}{x^+}\right)\ ,
}
which satisfy the constraint \eqref{con2}. It is important to understand the nature of the parameters above. The parameters $g$, $\alpha$, and of course $q$, are constants whereas $x^\pm$ and $\gamma$, of which two are independent due to the constraint \eqref{xll}, are kinematic variables that can vary on each one particle state. In \cite{Beisert:2008tw} Beisert and Koroteev identify a choice of $\gamma$ that has nice analytic properties. Introducing an arbitrary non-kinematic phase $\vpo$ that will be useful for discussing reality conditions in the later parts of this section, this choice of $\gamma$ is given by\footnote{As $\gamma$ can be understood as parametrising the normalisation of the fermionic states relative to the bosonic states the phase $e^{i\varphi_1}$ is not be physical as it can always be incorporated into the definition of the states. In the latter sections of this paper, we will take $\vpo =0$, whereas in \cite{Hoare:2011fj} it was taken to be equal to $\frac{\pi}{4}$.}
\EQ{
\gamma=e^{i\vpo}\frac{\sqrt{-i\alpha q^{C+1/2}U(x^+-x^-)}}{(1-(q-q^{-1})^2g^2)^{1/4}}\ ,
 \label{rev}
 }
 where $U=(x^+/qx^-)^{1/2}$
so that there is only a single kinematic variable which will be identified with the momentum of a one particle state.

At this point, we focus on two particular representations of $U_q(\mg)$ that are associated to the magnons and the solitons that are obtained as particular limits of the parameterization above. The  
magnon representation has been very well studied in the context of the string theory on $\text{\text{AdS}}_5\times S^5$ or $\mathcal{N}=4$ super Yang-Mills \cite{Beisert:2005tm,Beisert:2006qh,Klose:2006zd,Arutyunov:2006yd} and as such we just discuss the limit briefly. The soliton representation, however, is new and in the rest of this paper we will investigate this representation and its associated R-matrix.

{\bf Magnons:} This representation is constructed 
by taking the limit  of vanishing $q$-deformation, that is $q\to1$, or $k\to\infty$. The combination 
\EQ{
\frac{x^+}{x^-}=e^{ip}\ ,
\label{poy}
} 
where $p$, the kinematic variable, is the world sheet momentum of the string. The constant $g$ is a coupling which in the $\text{\text{AdS}}_5\times S^5$ setting is related to the 't~Hooft coupling by $g^2=\lambda/8\pi$. In the limit, the variable $\gamma$ is determined via \eqref{rev} with $\vpo=0$ to be
\EQ{
\gamma=\sqrt{-i\alpha e^{ip/2}(x^+-x^-)}\ .
}

{\bf Solitons:} This representation is obtained by keeping $q$ fixed and taking the limit $g\to\infty$, which is the limit of large string tension (or 't~Hooft coupling).  In this case, in contrast to \eqref{poy},
\EQ{
\frac{x^+}{x^-}=q\ .
}
First of all, taking the limit $g\to\infty$ of \eqref{rev} gives
\EQ{
\gamma=e^{i\vpo}\sqrt{\frac{\alpha x^+[\tfrac12]_q}{g}}\ ,\qquad [\tfrac12]_q=\frac1{q^{1/2}+q^{-1/2}}=\frac1{2\cos\tfrac\pi{2k}}
}
and then to leading order in $g^{-1}$
\EQ{
a=e^{i\vpo}\sqrt{\alpha x^+[\tfrac12]_q}\ ,\quad b=ie^{-2i\vpo}q^{-1/2}a\ ,\quad c=ie^{2i\vpo}q^{1/2}d\ ,\quad d=e^{-i\vpo}\sqrt{\frac{[\tfrac12]_q}{\alpha x^+}}\ .
\label{ma1} 
}
This implies that the central term $C=0$ and so the condition \eqref{con2} becomes
\EQ{
-PK=[\tfrac12]_q^2\ .
\label{con3}
}
This will be identified as a relativistic mass shell condition 
with $P$ and $K$ proportional to the lightcone components of the 2-momentum. We can define the kinematic variable $\theta$, to be identified with the rapidity, and take
\EQ{
x^\pm=\frac{q^{\pm1/2}}{\alpha}e^{-\theta +i\vpt}\ .
\label{ma2} 
}
In the above we have introduced a second arbitrary phase $e^{i\varphi_2}$.\footnote{In the later sections of this paper we take $\varphi_2 = 0$, whereas in \cite{Hoare:2011fj} it was taken to be $-\frac{\pi}{2}$.} With this parameterization
\EQ{
P=ab=i[\tfrac12]_q\, e^{-\theta + i\vpt}\ ,\qquad K=cd=i[\tfrac12]_q\, e^{\theta-i\vpt}\ .
}
In our interpretation, $\theta$ is the rapidity of the state, and $P$ and $K$ are proportional to the lightcone components 
\EQ{
p_\pm=\mu\sin\left(\frac{\pi}{2k}\right)e^{\pm\theta}
}
of the 2-momentum via
\EQ{
P=\frac{ip_-e^{i\vpt}}{\mu\sin\frac{\pi}k} ,\qquad K=\frac{ip_+e^{-i\vpt}}{\mu\sin\frac\pi k}\ .
\label{haa2} 
}
The constraint \eqref{con3}, is then interpreted as the mass-shell condition as promised:
\EQ{
p_+p_-=\mu^2\sin^2\left(\frac{\pi}{2k}\right)\ .
}
Notice that just as described in section \ref{s2} the rapidity appears in precisely the way dictated by the $\mathbb Z$ gradation \eqref{zgr} with Lorentz spin equal to minus half the $\mathbb Z$ grade. At the moment, it still is not obvious that we can associate $P$ and $K$ with the relativistic 2-momentum. The consistency of this identification will come when we consider the action of $\mP$ and $\mK$ on multi-particle states: the action will have the required additive property. 

In the soliton representation, using~\eqref{ma1}, the $\mathbb Z_4$ automorphism \eqref{dwe} can be written in the following way acting on the generators of the 4-dimensional representation of the quantum group
\EQ{
\sigma(M) = -{\cal K} M^{st} {\cal K}^{-1}\,,\label{z4aut}
}
where the super-transpose (with $M$ as in \eqref{vxx}) and ${\cal K}$ are defined as
\EQ{
M^{st}=\left(\begin{array}{c|c}m^t & -\eta^t \\\hline  \theta^t & n^t\end{array}\right)\,,\quad
{\cal K}=\left(\begin{array}{c|c} q^{-1/4}J & 0 \\\hline 0&-e^{2i\vpo}q^{1/4}J\end{array}\right)\ ,\quad J=\left(\begin{array}{cc} 0 & 1\\ -1 & 0\end{array}\right)\ .
} 
In the 
limit $q\to1$, with $\vpo = \frac\pi2$, the $\mathbb Z_4$ automorphism here becomes exactly the one used to define the semi-symmetric space \eqref{sss1}.
We remark that, in this limit, $\sigma$ is an outer automorphism of order 4 in the group of all automorphisms but, since $\sigma^2$ is inner, it has order 2 in the group of 
outer automorphisms~\cite{Arutyunov:2009ga,Serg3}. 

The soliton representation has the reality properties $a^*=-i\,e^{-i\vpt}\,b$, $d^*=-i\,e^{i\vpt}\,c$ so that
\ALAT{2}{
\notag({\mR}^a{}_b)^\dagger&={\mR}^b{}_a\ ,\qquad&({\mL}^\alpha{}_\beta)^\dagger&={\mL}^\beta{}_\alpha\ ,\\\label{rcpr}(\mQ^\alpha{}_b)^\dagger&=ie^{-i\vpt}\,\epsilon_{\alpha\beta}\epsilon^{ba}\,\mQ^\beta{}_a\ ,\qquad&
(\mS^a{}_\beta)^\dagger&=ie^{i\vpt}\,\epsilon_{ab}\epsilon^{\beta\alpha}\,\mS^b{}_\alpha\ ,\\
\notag{\mP}^\dagger&=-e^{-2i\vpt}\,{\mP}\ ,\qquad&{\mK}^\dagger&=-e^{2i\vpt}\,{\mK}\ .
} 
These are different to the usual reality conditions taken for the magnon representation, for which $a=d^*$ and $b=c^*$, implying \cite{Arutyunov:2006yd}
\ALAT{2}{
\notag( {\mR}^a{}_b)^\dagger&= {\mR}^b{}_a\ ,\qquad&( {\mL}^\alpha{}_\beta)^\dagger&= {\mL}^\beta{}_\alpha\ ,\\ \label{rcgs}
( {\mQ}^\alpha{}_b)^\dagger&= {\mS}^b{}_\alpha\ ,\qquad&
( {\mS}^a{}_\beta)^\dagger&= {\mQ}^\beta{}_a\ , \\
\notag {\mC}^\dagger&= {\mC}\ ,\qquad& {\mP}^\dagger&= {\mK}\ .
}

Of course, there will always exist an $SL(2,\mathbb C)$ automorphism that
relates the magnon and soliton representations that amounts to a
change in the basis of the generators of the algebra. One may think
that this allows a Lorentz symmetry to be defined on the magnon
representation. This is discussed further in the appendix, however, it appears that this is not consistent as the representations are not equivalent when one considers the way that they act on tensor product representations, a subject to which we now turn. At this point and for the rest of the paper we choose the arbitrary phases $\vpo$ and $\vpt$ to vanish. 

\subsection{The co-product}

The action of the quantum group on a tensor product in the Lie superalgebra case involves the co-product which generalizes \eqref{cop}
\ALAT{2}{
\notag\Delta(\mathfrak E_j)&=\mathfrak E_j\otimes 1+q^{-\mathfrak H_j}\mathfrak U^{s_j}\otimes \mathfrak E_j\ ,\qquad& 
\Delta(\mathfrak F_j)&=\mathfrak F_j\otimes q^{\mathfrak H_j}+\mathfrak U^{-s_j}\otimes\mathfrak F_j\ ,\\
\label{coprod}
\Delta(\mC)&=\mC\otimes 1+1\otimes\mC\ ,\qquad &
\Delta(\mP)&=\mP\otimes 1+q^{2\mC}\mU^2\otimes\mP\ ,\\
\notag\Delta(\mK)&=\mK\otimes q^{-2\mC}+\mU^{-2}\otimes\mK\ ,\qquad &
\Delta(\mU)&=\mU\otimes\mU\ .
\label{coprod} 
}
It involves a new abelian generator $\mU$ introduced in the magnon example to describe non-localities in the action of the supersymmetry generators on two-particle states. 

For consistency the coproduct for the central extensions should equal themselves under the action of the permutation operator on the tensor product. This imposes the following constraints  \cite{Beisert:2008tw}
\EQ{
\mP=g\alpha(1-q^{2\mC}\mU^2)\ ,\qquad\mK=g\alpha^{-1}(q^{-2\mC}-\mU^{-2})\ .
}
The braiding factor satisfies $\mU^2=x^+/qx^-\cdot{\bf1}$ with
\EQ{
\mU\ket{\phi^a}=\sqrt{\frac{x^+}{qx^-}}\ket{\phi^a}\ ,\qquad\mU\ket{\psi^\alpha}=
-\sqrt{\frac{x^+}{q x^-}}\ket{\psi^\alpha}\ .
}
In the magnon representation, from \eqref{poy}, $\mU$ acts as $e^{\mp ip/2}$ on states, while in the soliton representation
the braiding factor simplifies to
\EQ{
\mU\ket{\phi^a}=\ket{\phi^a}\ ,\qquad\mU\ket{\psi^\alpha}=
-\ket{\psi^\alpha}\ ,
}
which is just the fermion number. This is a significant result and required for the interpretation of the symmetry structure as the supersymmetry algebra of a relativistic QFT since, on physical grounds, one requires a factor $-1$ when moving a supersymmetry, including $\mathfrak E_2$ and $\mathfrak F_2$, past a fermionic state.\footnote{Note that in previous disucssions \cite{Beisert:2006qh,Arutyunov:2006yd,Beisert:2008tw} this minus sign was put in by hand and $\mU$ had the same eigenvalue on all states. It seems particularly nice that the minus sign can be incorporated into the definition of $\mU$.} 
Of crucial significance also is that in the soliton representation the non-trivial centres have a trivial co-product:
\EQ{
\Delta(\mP)=\mP\otimes 1+1\otimes\mP\ ,\qquad\Delta(\mK)=\mK\otimes 1+1\otimes\mK\ ,
}
which is required if we are to interpret them as the lightcone components of the 2-momentum.

The relation between the coproducts of the magnon and soliton representations is discussed more fully in the appendix.

\section{Representation Theory}\label{s4}

The representation theory of Lie superalgebras is much more convoluted than conventional Lie algebras. For a Lie algebra, arbitrary irreducible representations can be built up by taking tensor products of a small set of basic representations and decomposing. On the other hand, for Lie superalgebras, such tensor products are generally reducible but indecomposable. This feature, in particular, will play a prominent r\^ole in our story because physically the basic particles transform in the 4-dimensional representation of $\mg=\mpsl(2|2)\ltimes\mathbb R^3$ (or rather a tensor product of two copies thereof) and bound states are in representations that lie in tensor products of this representation. 

A further complication is that our algebra is $q$ deformed and this modifies the representation theory. Since we lack a comprehensive analysis of the representation theory of $U_q(\mg)$ we will make certain assumptions (although see \cite{Ky:1994cr,Ky:1994we}). We will take $q$ to be a generic deformation, in particular we will assume that $q$ is not a root of unity (a situation we shall analyse later). We will also assume that, as in the case of the $q$ deformation of an ordinary Lie algebra, the representations of $U_q(\mg)$ are simply deformations of the representations of $\mg$. This is supported by the explicit constructions of low-dimensional representations.

To start with we consider the undeformed algebra $\mg=\mathfrak{psl}(2|2)\ltimes{\mathbb R}^3$. As explained in section~\ref{s3}, 
 we can construct representations of this algebra by considering the analogous problem in the Lie superalgebra $\mathfrak{sl}(2|2)$ which has a single centre. The three centres can then be generated by the outer-automorphism group. 

\subsection{The long and the short representations}

Arbitrary representations of the related algebra $\mathfrak{gl}(2|2)$ were constructed 
in \cite{Zhang:2004qx} (see also~\cite{Beisert:2006qh,Beisert:2008tw}).
The algebra $\mathfrak{gl}(2|2)$ consists of the algebra $\msl(2|2)$ plus the additional generator ${\cal F}=\Be_{11}+\Be_{22}-\Be_{33}-\Be_{44}$ which plays the r\^ole of the fermion number. The simplest set of representations are of dimension $16(2J_1+1)(2J_2+1)$ and are labelled $(J_1,J_2,\mq,\mpf)$, where $J_i$ are $\msl(2)$ spins. Here, $\mq$ (not a deformation parameter) is identified with the eigenvalue of the central element, 
which in the defining representation is the identity matrix ${\bf1}$, and $\mpf$ is the fermion number label;
$\mq$ and $\mpf$ are complex numbers. If we ignore the fermion label $\mpf$ then these representations give representations of $\mathfrak{sl}(2|2)$. These are the {\it long\/}, or {\it typical\/}, representations denoted $\{m,n\}$, with $m=2J_1$ and $n=2J_2$, in \cite{Beisert:2008tw}. These representations exist for generic values of the single centre $C=\mq$ and by making use of the $\msl(2,\C)$ 
automorphism we can use them to construct representation of the case with general values for all 3 centres with $C^2-PK=\mq^2$.

When the centre $\mq$ takes special values the long representations become reducible but indecomposable. This is called a {\it shortening\/} condition and it is very similar to a BPS condition in a supersymmetric QFT.
What happens is that the corresponding module $V_{\{m,n\}}$ splits as
\EQ{
V_{\{m,n\}}=V_\text{sub-rep}\oplus V^\perp\ ,
\label{kjj}
}
where $V_\text{sub-rep}$ is an invariant subspace under the action of $\mg$. This is therefore a representation of $\mg$, the {\it sub-representation\/}. What makes $\{m,n\}$ indecomposable is that $V^\perp$ is {\it not\/} an invariant subspace. However, the quotient
\EQ{
V_\text{factor}=\coset{V_{\{m,n\}}}{V_\text{sub-rep}}
}
defines another representation of $\mg$, the {\it factor representation\/}. In a basis for the module $\MAT{u\\ v}$, $u\in V_\text{sub-rep}$ and $v\in V^\perp$, when the shortening condition holds,
the generators of the algebra take the form
\EQ{
\MAT{* & *\\ 0 & *}\ .
} 

The sub- and factor representations are known as short, or atypical, representations.
There are four possibilities that we consider below~\cite{Zhang:2004qx}:

(i) $\mq=\tfrac12(m+n+2)$. In this case, the sub-representation has dimension 
\EQ{
4(2mn+3m+n+2)
}
and we denote it 
$\langle m,n+1\rangle$ to agree with the notation of \cite{Beisert:2008tw}. The corresponding factor representation is then $\langle m+1,n\rangle$ and has dimension 
\EQ{
4(2mn+m+3n+2)\ .
}

(ii)  $\mq=-\tfrac12(m+n+2)$. In this case the situation is the same as (i) except that the r\^oles of the sub and factor representations are interchanged.

(iii) $\mq=\tfrac12(m-n)$, $m\neq n$.
In this case, the sub-representation has dimension 
\EQ{
4(2mn+m+n)
}
and we denote it $\langle m,n\rangle_2$. The corresponding factor representation has dimension
\EQ{
4(2mn+3m+3n+4)\ .
}
and we denote it as $\langle m,n\rangle_3$.

(iv) $\mq=-\tfrac12(m-n)$, $m \neq n$. 
In this case the situation is the same as (i) except that the r\^oles of the sub and factor representations are interchanged.

(v) $\mq = 0$, $m = n \neq 0$. 
In this case, the sub-representation has dimension
\EQ{
2(2m^2+4m+1)
}
and we denote it $\langle m,m \rangle_4$. The corresponding factor representation has dimension
\EQ{
2(6m^2+12m+7)\,.
}

(vi) $\mq = 0$, $m = n = 0$. For this special case we have that the subrepresentation is a singlet, denoted by $\bullet$, and
\EQ{
\{0,0\}\longrightarrow\bullet\oplus\text{\bf adj}\oplus\bullet\ .
}

For our purposes, we will be mostly interested in the atypical representations
$\langle m,n\rangle$ whose dimension is $4(m+1)(n+1)+4mn$. For later use, the
representations $\langle m,0\rangle$ have dimension $4(m+1)$, $\mq=\frac{1}{2}(m+1)$, and $\mg_0$ content\,\footnote{That is, this is the decomposition into representations of the zero graded part of the algebra $\mg_0=\mathfrak{sl}(2)\oplus\mathfrak{sl}(2)$.} 
\EQ{
\langle m,0\rangle=(m+1,0)\oplus(m,1)\oplus(m-1,0)\ .
\label{juu}
}
The 4-dimensional defining representation corresponds to $\langle0,0\rangle$.

For the algebra with 3 non-vanishing centres, the shortening conditions (i) and (ii) can be written
\EQ{
\mq^2=C^2-PK=\tfrac14(m+n+2)^2:\qquad\{m,n\}\longrightarrow\langle m,n+1\rangle+\langle m+1,n\rangle\ ,
\label{er2}
}
where one of the representations on the right-hand side is the sub-representation and one the factor representation. 
In the similar way, for $m \neq n$ the shortening conditions (iii), (iv) 
are
\EQ{
\mq^2=C^2-PK=\tfrac14(m-n)^2:\qquad\{m,n\}\longrightarrow\langle m,n\rangle_2+\langle m,n\rangle_3\ .
\label{er1}
}

When we move to the quantum algebra $U_q(\mg)$ the shortening conditions on $\{m,n\}$ in the above must become suitably deformed. Following, and generalizing \cite{Beisert:2008tw}, we propose that the conditions \eqref{er2} and \eqref{er1} become
\EQ{
[C]_q^2-PK=[\tfrac12(m+n+2)]_q^2\ ,\qquad
[C]_q^2-PK=[\tfrac12(m-n)]_q^2\ ,\label{sh2}
}
respectively.
This can be checked for small values of $m$ and $n$ explicitly and we shall assume that it is true generally. In the deformed theory, the atypical representations $\langle m,n\rangle$ require
\EQ{
[C]_q^2-PK=[\tfrac12(m+n+1)]_q^2\ .
\label{bps}
}

As we reported in section \ref{s2}, an 
S-matrix theory constructed from the quantum group requires a perfect meshing of the representation theory with the analytic structure and in this regard the shortening conditions \eqref{sh2} will play a key r\^ole. 
For the construction of the S-matrix, we will be particularly interested in the representations $\{m,0\}$ of $U_q(\mg)$. For these representations, the first shortening condition in \eqref{sh2} corresponds to
\EQ{
\{m,0\}\longrightarrow\langle m+1,0\rangle\oplus\langle m,1\rangle\ ,
}
while for $m>0$ 
the second condition in \eqref{sh2} corresponds to
\EQ{
\{m,0\}\longrightarrow\langle m,0\rangle_2\oplus\langle m,0\rangle_3\ . 
}
Note that $\langle m,0\rangle_2\equiv\langle m-1,0\rangle$ for $m>0$. The special case where $m=0$ and the shortening condition $\eqref{sh2}$ is satisfied is given by case (vi) above. 

The other important information we need, is the decomposition of a tensor product of the particular short representations $\langle m,0\rangle$, $m\geq0$. These take the form
\EQ{
\langle m,0\rangle\otimes \langle n,0\rangle= \sum_{k=0}^{\text{min}(m,n)} \{m+n-2k,0\}\ .
\label{xxy}
}

\section{The Basic S-Matrix}\label{s5} 

For our relativistic QFT, we will be identifying the particle states with the short representations $\pi^\theta_a=\langle a-1,0\rangle$ with associated modules $V_a(\theta)$, $a\in\mathbb N$. 
The masses of the states follow from the shortening, or BPS, condition \eqref{bps} with $C=0$ and $P$ and $K$ related to the lightcone components of the momentum as in \eqref{haa2}: 
\EQ{
m_a=\mu\sin\left(\frac{\pi a}{2k}\right)\ ,
\label{mass}
}
which is the mass formula \eqref{mass2}.
Obviously, this formula suggests that $a$ should somehow be cut-off appropriately, an issue we will return to later. In particular, 
the basic states transform in the 4-dimensional representation $\pi_1^\theta$. The 2-body S-matrix elements of the basic particles involves the tensor product
\EQ{
\pi^{\theta_1\theta_2}_{11}=\langle 0,0\rangle\otimes\langle 0,0\rangle\ .
}

The $\check R$-matrix for the tensor product $V_1\otimes V_1$ can be extracted from the general solution in \cite{Beisert:2008tw} by taking the limit $g\to\infty$ \cite{Beisert:2010kk,Hoare:2011fj} 
and matching the parameters as in \eqref{ma1} and \eqref{ma2}. The explicit expression for the $\check R$-matrix in the basis $\{\ket{\phi^a},\ket{\psi^\alpha}\}$, with $x=e^{\theta_{12}}$, is
\EQ{ 
\check R(x)\ket{\phi^a\phi^a}&=A\ket{\phi^a\phi^a}\ ,\qquad \check R(x)\ket{\psi^\alpha\psi^\alpha}=D\ket{\psi^\alpha\psi^\alpha}\ ,\\
\check R(x)\ket{\phi^1\phi^2}&=\frac{q(A-B)}{q^2+1}\ket{\phi^2\phi^1}+
\frac{q^2 A+B}{q^2+1}\ket{\phi^1\phi^2}+\frac{C}{1+q^2}\ket{\psi^1\psi^2}-\frac{qC}{1+q^2}\ket{\psi^2\psi^1}\ ,\\
\check R(x)\ket{\phi^2\phi^1}&=\frac{q(A-B)}{q^2+1}\ket{\phi^1\phi^2}
+\frac{q^2 B+A}{q^2+1}\ket{\phi^2\phi^1}-\frac{qC}{1+q^2}\ket{\psi^1\psi^2}+\frac{q^2C}{1+q^2}\ket{\psi^2\psi^1}\ ,\\
\check R(x)\ket{\psi^1\psi^2}&=\frac{q(D-E)}{q^2+1}\ket{\psi^2\psi^1}
+\frac{q^2 D+E}{q^2+1}\ket{\psi^1\psi^2}+\frac{F}{1+q^2}\ket{\phi^1\phi^2}-\frac{qF}{1+q^2}\ket{\phi^2\phi^1}\ ,\\
\check R(x)\ket{\psi^2\psi^1}&=\frac{q(D-E)}{q^2+1}\ket{\psi^1\psi^2}
+\frac{q^2 E+D}{q^2+1}\ket{\psi^2\psi^1}-\frac{qF}{1+q^2}\ket{\phi^1\phi^2}+\frac{q^2F}{1+q^2}\ket{\phi^2\phi^1}\ ,\\
\check R(x)\ket{\phi^a\psi^\alpha}&=G\ket{\psi^\alpha\phi^a}+H\ket{\phi^a\psi^\alpha}\ ,\qquad
\check R(x)\ket{\psi^\alpha\phi^a}=K\ket{\psi^\alpha\phi^a}+L\ket{\phi^a\psi^\alpha}\ ,
}
with\footnote{Our functions are those of \cite{Beisert:2008tw} multiplied by $(x-q)(x+1)/(q^{1/2}x)$ in order to ensure that $\check R(x)$ has no poles. Similarly compared to those in \cite{Hoare:2011fj} we have multiplied the functions by $x-x^{-1}$ and rescaled the fermionic states by a factor of $e^{-i\pi/4}$. This is related to the choice of $\gamma$ (which controls the normalisation of fermions relative to bosons) \eqref{ma2} which differs from that of \cite{Hoare:2011fj} by precisely this factor. Also recall that in \cite{Hoare:2011fj} $q$ was taken to be related to $k$ as $q=e^{-i\pi/k}$, see footnote \ref{footnote1}.
}
\ALAT{2}{ 
\notag A&=\frac{(q x-1)(x+1)}{q^{1/2}x}\ ,\qquad& D&=\frac{(q-x)(x+1)}{q^{1/2}x}\ ,\\ 
\notag B&=\frac{q^3-(q^3-2q^2+2q-1)x-x^2}{q^{3/2}x} \ ,\qquad& E&=\frac{q^3x^2-(q^3-2q^2+2q-1)x-1}{q^{3/2}x}\ , \\
\notag C&=F=\frac{i(q-1)(q^2+1)(x-1)}{q^{3/2} x^{1/2}}\  ,\qquad& G&=L=x-x^{-1}\ ,\\ H&=K=\frac{(q-1)(x+1)}{q^{1/2}x^{1/2}}\ .
}
Since the tensor product has $U_q(\mg_0)$ content
\EQ{
\langle0,0\rangle\otimes\langle0,0\rangle=(2,0)\oplus(0,2)\oplus2(1,1)\oplus2(0,0)\ ,
\label{dcp}
}
another way to express the $\check R$-matrix is in terms of $U_q(\mg_0)$ invariant projectors
\EQ{
\check R(x)&=\frac{(x+1)(q-x)}{x\sqrt q}\Pr_{(0,2)}
+\frac{(x+1)(qx-1)}{x\sqrt q}\Pr_{(2,0)}\\ &+
\frac{(x+1)(\sqrt{qx}-1)(\sqrt q+\sqrt x)}{x\sqrt q}\Pr_{(1,1)}^{(+)}+
\frac{(x+1)(\sqrt{qx}+1)(\sqrt q-\sqrt x)}{x\sqrt q}\Pr_{(1,1)}^{(-)}\\ &+
f_+(x)\Pr_{(0,0)}^{(+)}+f_-(x)\Pr_{(0,0)}^{(-)}\ ,
\label{srm}
}
where
\EQ{
f_\pm(x)&=\frac1{2q^{3/2}x^2}\Big(
q^3(x-1)^2+4qx(q-1)
\pm(x-1)\Big(1-x+\big((x-1)^2+q^6(x-1)^2\\ &\qquad\qquad\qquad
+4xq(2q^4-3q^3-3q+2)+2q^3(1+10x+x^2)\big)^{1/2}\Big)\Big)\ .
}

In order to construct an S-matrix we will need the unitarity condition
\EQ{
\check R(x^{-1})\check R(x)=\frac{(q-x)(qx-1)(x+1)^2}{qx^2}1\otimes 1
}
and the crossing symmetry relation
\EQ{
\check R(x)=({\cal C}^{-1}\otimes1)(\sigma\cdot
\check R(-x^{-1}))^{st_1}\cdot\sigma\cdot(1\otimes{\cal C})\ ,
}
where $\sigma$ is the graded permutation operator and $st$ indicates the super-transpose on first space in the tensor product.\footnote{Note that $\sigma\cdot\check R(-x^{-1})\in\text{End}\,(V_1\otimes V_1)$ so the transpose is well defined. In terms of explicit indices $(A^{st})_{ab}=A_{ba}$, $(A^{st})_{a\alpha}=-A_{\alpha a}$, $(A^{st})_{\alpha a}=A_{a\alpha}$ and $(A^{st})_{\alpha\beta}=A_{\beta\alpha}$.}
The charge conjugation matrix on the basic states takes the form
\ALAT{2}{
\notag{\cal C}\ket{\phi^1}&=q^{-1/2}\ket{\phi^2}\ ,\qquad&{\cal C}\ket{\phi^2}&=-q^{1/2}\ket{\phi^1}\ ,\\
\quad{\cal C}\ket{\psi^1}&=q^{-1/2}\ket{\psi^2}\ ,\qquad&{\cal C}\ket{\psi^2}&=-q^{1/2}\ket{\psi^1}\ . 
} 
In order that the S-matrix satisfies the unitarity and crossing symmetry constraints,  \eqref{unit} and \eqref{cs}, we multiply the $\check R$-matrix by a scalar function
\EQ{
\widetilde S_{11}(\theta)=Y(\theta)Y(i\pi-\theta)\check R(e^{\theta})\ .
}
This form guarantees that crossing symmetry is satisfied and unitarity requires
\EQ{
Y(\theta)Y(i\pi-\theta)Y(-\theta)Y(i\pi+\theta)=\frac{1}{16\sinh(\frac\theta{2}+\frac{i\pi}{2k})
\sinh(-\frac\theta{2}+\frac{i\pi}{2k})\cosh^2(\frac\theta{2})}\ .
}
The solution to this is not unique, however, there is a minimal solution with the smallest number of poles and zeros and, in particular, with no poles on the physical strip; namely,
\EQ{
Y(\theta)=\frac1{\sqrt2\pi}
\prod_{l=0}^\infty\frac{\Gamma(\tfrac{\theta}{2i\pi}+l+\tfrac1{2k})\Gamma(\tfrac{\theta}{2i\pi}+l-\tfrac1{2k}+1)}{\Gamma(\tfrac{\theta}{2i\pi}+l+\tfrac1{2k}+\tfrac12)\Gamma(\tfrac{\theta}{2i\pi}+l-\tfrac1{2k}+\tfrac32)}\cdot\frac{\Gamma(\tfrac{\theta}{2i\pi}+l+\tfrac1{2})^2}{\Gamma(\tfrac{\theta}{2i\pi}+l+1)^2}\ .
}
We can also write the integral representation
\EQ{
&Y(\theta)Y(i\pi-\theta)
=\frac{{\cal F}(\theta)}{2(q-q^{-1})}\ ;\\
&{\cal F}(\theta)=
\exp\left[-2\int_0^\infty\frac{dt}t
\,\frac{\cosh^2(t(1-\tfrac1k))\sinh(t(1-\tfrac\theta{i\pi}))
\sinh(\tfrac{t\theta}{i\pi})}{\sinh t\cosh^2t}\right]\ .
\label{gqq}
}
Notice that ${\cal F}(\theta)$ is real and positive when $\theta$ is purely imaginary. 

\section{The Bootstrap Programme}\label{s6} 

As in the 
non-graded affine Lie algebra case described in section \ref{s2}, the representation $\pi^{\theta_1\theta_2}_{11}$ is irreducible for generic values of $\theta_1$ and $\theta_2$. In fact it is identified with the 16-dimensional long representation $\{0,0\}$. However, drawing on the results reported earlier in section \ref{s4}, 
for special values of the rapidity difference $\theta_{12}$ the representation becomes reducible. Setting $C_1=C_2=0$, the tensor product has $P=P_1+P_2$ and $K=K_1+K_2$ and so the first shortening condition in \eqref{sh2} becomes
\EQ{
-(P_1+P_2)(K_1+K_2)=[\tfrac32]_q^2\qquad\implies\qquad
\theta_{12}=\pm\frac{i\pi}k\ .
\label{ewe}
}
At these special points, the representation becomes reducible
\EQ{
\{0,0\}\longrightarrow\langle1,0\rangle\oplus\langle0,1\rangle
}
and for the upper sign $\langle0,1\rangle$ and $\langle1,0\rangle$ are the sub- and factor  representation, respectively. For the lower sign these designations swap over. 

At these points the $\check R$-matrix, and hence the S-matrix gains a non-trivial kernel. The fact that $\widetilde S_{11}(\theta)$ lies in the commutant of $U_q(\mg)$ means that, for consistency, the kernel must be the invariant subspace corresponding to the sub-representation. The bound state is consequently associated to the factor representation in the tensor product. At this point we have a choice to make. By picking the sign of $k$, we can choose either special point to be on the physical strip. Here, we will take $k$ to be positive, in which case the special point $\theta_{12}=\frac{i\pi}k$ lies on the physical strip and the potential bound state corresponds to the representation $V_2=\langle1,0\rangle$. In this case, the kernel of $\widetilde S_{11}(\tfrac{i\pi}k)$ corresponds to the sub-reprentation $\langle0,1\rangle$:
\EQ{
\widetilde S_{11}(\tfrac{i\pi}k):\quad V_{\langle0,1\rangle}\longrightarrow0\ .
}
The bound state transforms in the factor representation and we write
\EQ{
\pi^\theta_2\subset\pi^{\theta+\tfrac{i\pi}{2k},\theta-\tfrac{i\pi}{2k}}_{11}\Big|_\text{factor}\ .
}
On the other hand, because $\widetilde S_{11}(\frac{i\pi}k)$ permutes the rapidities it swops over the special points \eqref{ewe} and so maps the factor representation $\langle1,0\rangle\subset\{0,0\}$ to the sub-representation $\langle1,0\rangle\subset\{0,0\}$, and we write
\EQ{
\pi^\theta_2\subset\pi^{\theta-\tfrac{i\pi}{2k},\theta+\tfrac{i\pi}{2k}}_{11}\Big|_\text{sub}\ .
}

Since the whole issue of the shortening of the tensor product representation is key to the construction of the S-matrix, we will pause to discuss it in explicit detail.
The tensor product module $V_1\otimes V_1$ can be decomposed in terms of $U_q(\mg_0)$ modules, following the decomposition \eqref{dcp}, as:
\EQ{
V^{\Sp}_{\{0,0\}}=V^{\Sp}_{(2,0)}\oplus V_{(1,1)}^{(+)}\oplus V_{(1,1)}^{(-)}\oplus V^{\Sp}_{(0,2)}\oplus V_{(0,0)}^{(+)}\oplus V_{(0,0)}^{(-)}\ .} 
Explicitly, we have the bases\footnote{The choice of the singlets is made so that the invariant subspaces that we define in due course are simpler.}
\EQ{
V^{\Sp}_{(2,0)}:\quad&\ket{\phi^1\phi^1}\ ,\quad q^{1/2}\ket{\phi^1\phi^2}+q^{-1/2}\ket{\phi^2\phi^1}\ ,\quad\ket{\phi^2\phi^2}\ ,\\
V^{(\pm)}_{(1,1)}:\quad&\ket{\phi^1\psi^1}\pm\ket{\psi^1\phi^1}\ ,\quad\ket{\phi^2\psi^1}\pm\ket{\psi^1\phi^2}\ ,\\ &\ket{\phi^1\psi^2}\pm\ket{\psi^2\phi^1}\ ,\quad
\ket{\phi^2\psi^2}\pm\ket{\psi^2\phi^2}\ ,\\
V^{\Sp}_{(0,2)}: \quad &\ket{\psi^2\psi^2}\ ,\quad q^{1/2}\ket{\psi^1\psi^2}+q^{-1/2}\ket{\psi^2\psi^1}\ ,\quad\ket{\psi^1\psi^1}\ ,\\
V^{(+)}_{(0,0)}: \quad& \frac{i(q^{1/2}-q^{-1/2})}{q+q^{-1}}\big(q^{-1/2}\ket{\phi^1\phi^2}- q^{1/2}\ket{\phi^2\phi^1}\big)
+q^{-1/2}\ket{\psi^1\psi^2}-q^{1/2}\ket{\psi^2\psi^1}\ ,\\
V^{(-)}_{(0,0)}: \quad& q^{-1/2}\ket{\phi^1\phi^2}- q^{1/2}\ket{\phi^2\phi^1}
-\frac{i(q^{1/2}-q^{-1/2})}{q+q^{-1}}\big(q^{-1/2}\ket{\psi^1\psi^2}-q^{1/2}\ket{\psi^2\psi^1}\big)\ .}

Generically, $\mE_2$ and $\mF_2$ act between the $U_q(\mg_0)$ modules according to figure~\ref{f6}.  
When $\theta_{12}=\frac{ i\pi}k$ the situation is shown in figure~\ref{f7} 
where the action along the dotted lines is in one direction only as indicated by the arrows. The subspace $V_{(0,2)}\oplus V_{(1,1)}^{(-)}\oplus V_{(0,0)}^{(-)}$ becomes an invariant subspace and forms a module for the sub-representation $\langle0,1\rangle$. When $\theta_{12}=-\frac{i\pi}k$ the arrows are reversed
\begin{figure}
\begin{center}
\begin{tikzpicture}[line width=1.5pt]
\matrix (m) [matrix of math nodes, row sep=3em, column sep=2.5em, text height=1.5ex, text depth=0.25ex] {  & V_{(1,1)}^{(+)} & V_{(0,0)}^{(-)} &  \\
V_{(2,0)}& & &  V_{(0,2)}\\ & V_{(0,0)}^{(+)}&V_{(1,1)}^{(-)}&\\ };
\path[<->]
(m-2-1) edge  (m-1-2);
\path[<->]
(m-1-2) edge  (m-1-3);
\path[<->]
(m-3-2) edge  (m-3-3);
\path[<->]
(m-3-3) edge  (m-2-4);
\path[<->]
(m-1-2) edge  (m-3-2);
\path[<->]
(m-1-3) edge  (m-3-3);
\path[<->]
(m-2-1) edge  (m-3-3);
\path[<->]
(m-1-2) edge  (m-2-4);
\end{tikzpicture} 
\end{center}
\caption{Action of the supersymmetry generators $\mathfrak E_2$ and $\mathfrak F_2$ on the $U_q(\mg_0)$ submodules of $V_1\otimes V_1$ at generic $\theta_{12}$.}
\label{f6}
\end{figure}
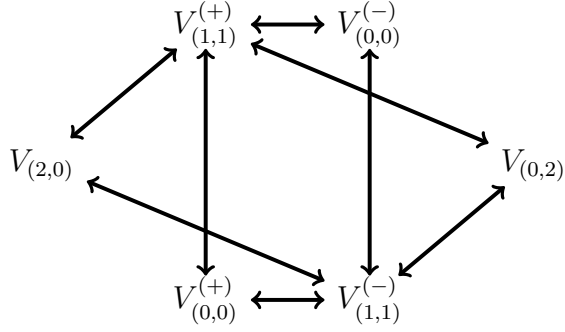
\begin{figure}
\begin{center}
\begin{tikzpicture}[line width=1.5pt]
\matrix (m) [matrix of math nodes, row sep=3em, column sep=2.5em, text height=1.5ex, text depth=0.25ex] {  & V_{(1,1)}^{(+)} & V_{(0,0)}^{(-)} &  \\
V_{(2,0)}& & &  V_{(0,2)}\\ & V_{(0,0)}^{(+)}&V_{(1,1)}^{(-)}&\\ };
\path[<->]
(m-2-1) edge  (m-1-2);
\path[->,dashed]
(m-1-2) edge  (m-1-3);
\path[->,dashed]
(m-3-2) edge  (m-3-3);
\path[<->]
(m-3-3) edge  (m-2-4);
\path[<->]
(m-1-2) edge  (m-3-2);
\path[<->]
(m-1-3) edge  (m-3-3);
\path[->,dashed]
(m-2-1) edge  (m-3-3);
\path[->,dashed]
(m-1-2) edge  (m-2-4);
\end{tikzpicture} 
\end{center}
\caption{Action of the supersymmetry generators $\mathfrak E_2$ and $\mathfrak F_2$ on the $U_q(\mg_0)$ submodules of $V_1\otimes V_1$ when $\theta_{12}=\frac{ i\pi}k$. The dotted lines have one-sided arrows.}
\label{f7}
\end{figure}
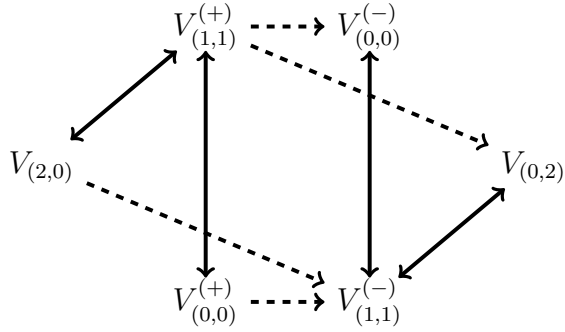

The S-matrix of the bound-state $V_2$ can then be found by using the bootstrap/fusion equations \eqref{bfs}. At the special rapidity difference, using \eqref{srm}, we have
\EQ{
\widetilde S_{11}(\tfrac{i\pi}k)&=\frac{{\cal F}(i\pi/k)}{2(q-q^{-1})}\check R(q)\\ &=
{\cal F}\left({i\pi}/k\right)\left[
\frac{q+1}{2q^{1/2}}{\mathbb P}_{(2,0)}+{\mathbb P}_{(1,1)}^{(+)}+
\frac{q^4-q^3+4q^2-q+1}{2q^{3/2}(q+1)}{\mathbb P}_{(0,0)}^{(+)}\right]\ .
\label{res}
}
which is non-vanishing on the factor representation $\langle1,0\rangle$ and vanishing on the sub-representation $\langle0,1\rangle$, as required for the consistency of the bootstrap \eqref{ipm}. Notice,
that the S-matrix residue is not a projector rather it is a weighted sum of $U_q(\mg_0)$ projectors; precisely the situation described in section~\ref{s2}. In the above, ${\cal F}(\frac{i\pi}k)$ is a postive real number given by the exponential in \eqref{gqq} evaluated at $\theta=\tfrac{i\pi}k$ and for $k\in{\mathbb R}>0$, the $\rho_j$ above are all real numbers --- in fact positive --- which is a necessary condition for the unitarity of the underlying QFT.\footnote{A more in-depth analysis will be needed to constrain the signs the residues.}

The bootstrap/fusion equations \eqref{bfs} can be used to write down the S-matrix for the scattering of $V_1$ with $V_2$
\EQ{
\widetilde S_{12}(\theta)=\left(1\otimes\widetilde S_{11}(\theta+\tfrac{i\pi}{2k})\right)\left(\widetilde S_{11}(\theta-\tfrac{i\pi}{2k})\otimes1\right)\Big|_{V_1\otimes V_2}\ ,
\label{fus1}
}
Note, that in the above, we have not shown the explicit projection factors present in \eqref{bfs} since in this case the decomposition of $V_2$ into $U_q(\mg_0)$ representations is non-degenerate and the S-matrix acts diagonally: $V_1\otimes V_2^{(j)}\to V_2^{(j)}\otimes V_1$. In that case, \eqref{bfs} is only non-vanishing when $j=l$ and so the $\rho_j$ factors are not needed. The projection onto $V_2$ is then only indicated implicitly.

The bootstrap programme now continues. The blueprint for the resulting theory is as follows. Particles are associated to the representations $\pi^\theta_a=\langle a-1,0\rangle$
with masses as given in \eqref{mass}. Each particle is self conjugate $a=\bar a$. The appearance of  bound states is governed by the three-point 
 couplings at rapidity angles
\EQ{
u_{ab}^{a+b}=\frac{\pi(a+b)}{2k}\ ,\qquad u_{ab}^{|a-b|}=\pi-\frac{\pi|a-b|}{2k} \label{ded}
}
illustrated in figure~\ref{f8}. 
Note, here, that the second is implied by the first and the fact that the particles are self conjugate. The scalar factor $X_{ab}(\theta)$ provides simple poles on the physical strip at these rapidity differences, as well as poles corresponding to bound states in the crossed channel at $\theta=i(\pi-u_{ab}^{a+b})$ and $\theta=i(\pi-u_{ab}^{|a-b|})$. 
\FIG{
\begin{tikzpicture} [line width=1.5pt,inner sep=2mm,
place/.style={circle,draw=blue!50,fill=blue!20,thick},proj/.style={circle,draw=red!50,fill=red!20,thick}]
\node at (2,2) [proj] (p1) {};
\node at (0.4,0.4) (i1) {$a$};
\node at (3.6,0.4) (i2)  {$b$};
\node at (2,4) (i3) {$a+b$};
\draw[-]  (i1) -- (p1);
\draw[-]  (i2) -- (p1);
\draw[-]  (i3) -- (p1);
\node at (2,0.3) {$\frac{\pi(a+b)}{2k}$};
\node at (8,2) [proj] (p1) {};
\node at (6.4,0.4) (i1) {$a$};
\node at (9.6,0.4) (i2)  {$b$};
\node at (8,4) (i3) {$|a-b|$};
\draw[-]  (i1) -- (p1);
\draw[-]  (i2) -- (p1);
\draw[-]  (i3) -- (p1);
\node at (8,0.3) {$\pi-\frac{\pi|a-b|}{2k}$};
\end{tikzpicture}
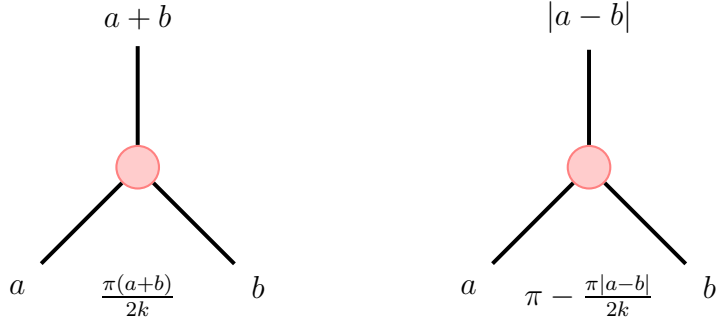
\caption{\small The 3-point couplings and associated rapidity angles. Note that both diagrams are equivalent since the particles are self-conjugate.}
\label{f8}
}

For instance the S-matrix element $S_{12}(\theta)$ has four simple poles. The poles at 
$\theta=\frac{3i\pi}{2k}$ and $\frac{i\pi}{2k}$ should correspond to particles $V_3$ and $V_1$, respectively, in the direct channel. The questions is does this mesh with the quantum group representation theory? The tensor product representation in question is
\EQ{
\pi^{\theta_1\theta_2}_{12}=\langle0,0\rangle\otimes\langle1,0\rangle\ .
} 
According to the representation theory of the undeformed algebra $\mg$, we expect this to be the irreducible representation $\{1,0\}$ for generic $\theta_{12}$. Now the simple pole at $\theta_{12}=\frac{3i\pi}{2k}$ occurs precisely at the first shortening condition in \eqref{sh2} of the representation $\{1,0\}$ corresponding to the factor representation $\langle2,0\rangle$. At this point we can verify directly that the S-matrix $\widetilde S_{12}$ 
that is constructed from the fusion equations \eqref{fus1} indeed is only non-vanishing on $\langle2,0\rangle$ which has $U_q(\mg_0)$ content $(3,0)\oplus(2,1)\oplus(1,0)$:
\EQ{
\widetilde S_{12}(\tfrac{3i\pi}{2k})&={\cal F}\left({2i\pi}/k\right){\cal F}\left({i\pi}/k\right)
\left[
\frac{(q^2+1)(q^2+q+1)}{4q^2}\Bbb P_{(3,0)}\right.\\ &\left.+\frac{(q^2+1)(q+\sqrt q+1)}{4q^{3/2}}\Bbb P_{(2,1)}+\frac{q^3+q^{5/2}+q^{2}+q^{1/2}+1}{4q^{3/2}}
\Bbb P_{(1,0)}\right]\ .\label{wts12a}  
}
This matches the fact that, from \eqref{ded}, particle 3 can be formed as a bound state of  1 and 2. The coefficients of the intertwiners in the above are all real and positive. 

However, the representation $\{1,0\}$ admits another shortening condition, the second in \eqref{sh2}, occurring on the physical strip at $\theta_{12}=i\pi-\frac{i\pi}{2k}$ corresponding to the representation $\langle0,0\rangle$. Once again we can verify directly that the S-matrix $\widetilde S_{12}$-matrix that is constructed from the fusion equations \eqref{fus1} is only non-vanishing on $\langle0,0\rangle$ which has $U_q(\mg_0)$ content $(1,0)\oplus(0,1)$:
\EQ{
\widetilde S_{12}(i\pi-\tfrac{i\pi}{2k})\propto i(q^{1/2}-q^{-1/2})\Bbb P_{(1,0)}+\big(q+q^{-1}\big)\Bbb P_{(0,1)}\label{wts12b} 
}
and this matches the other three-point coupling in \eqref{ded}.

The picture for general $S_{ab}(\theta)$ is now clear. The tensor product representation
\EQ{
\pi^{\theta_1\theta_2}_{ab}=\langle a-1,0\rangle\otimes\langle b-1,0\rangle
}
is, according to the decomposition for the undeformed algebra, the reducible representation \eqref{xxy}
\EQ{
\{a+b-2,0\}\oplus\{a+b-4\}\oplus\cdots\oplus\{|a-b|,0\}
}
the simple pole at $\theta_{12}=\frac{i\pi(a+b)}{2k}$ occurs precisely at the special point where $\{a+b-2,0\}$ becomes reducible with a factor representation $\langle a+b-1,0\rangle$. At this point, $\widetilde S_{ab}(\tfrac{i\pi(a+b)}{2k})$ is only non-vanishing on this subspace. Correspondingly, at the simple pole  $\theta_{12}=i\pi-\frac{i\pi|a-b|}{2k}$ the representation $\{|a-b|,0\}$ becomes reducible with a factor representation $\langle |a-b|-1,0\rangle$. At this point, $\widetilde S_{ab}(i\pi-\tfrac{i\pi|a-b|}{2k})$ is only non-vanishing on that subspace.

Although we have not proved the above picture for arbitrary $a$ and $b$, we have checked that the S-matrix has the required projection properties for the case $a=1$ and $b=3$.

\subsection{A magnon-like relativistic S-matrix}\label{s6.1}

The S-matrix building blocks $\widetilde S_{ab}(\theta)$ can then be put together with a scalar factor which supplies necessary poles on the physical strip. A magnon-like S-matrix is obtained by putting together two such blocks in a graded tensor product so that the symmetry algebra is a product
$U_q(\mg)\times U_q(\mg)$. In addition, the central elements are identified since each factor has the same momentum or rapidity.
Particles transform in product representations $\langle a-1,0\rangle\otimes_\text{gr}\langle a-1,0\rangle$.  The S-matrix elements take the form\,\footnote{One can check that the graded tensor product structure does not upset the Yang-Baxter Equation that is satisfied by each factor separately.}
\EQ{
S_{ab}(\theta)=X_{ab}(\theta)\,\widetilde S_{ab}(\theta)\otimes_\text{gr}\widetilde S_{ab}(\theta)\ ,\label{smfull}
}
where the tensor product is graded meaning that it respects boson/fermion statistics. The scalar factor $X_{ab}$ satisfies the bootstrap equations by itself and so is defined by specifying it on the basic particle $a=b=1$:\footnote{This choice for the scalar factor along with the phase factor \eqref{smfull} implies the full S-matrix for the scattering of two basic particles agrees with that constructed in \cite{Hoare:2011fj} up to the sign in $k$. The sign in $k$ amounts to the opposite choice of which states are bosonic and fermionic for the factorised S-matrix.}
\EQ{
X_{11}(\theta)=\frac{\sinh(\tfrac{\theta}{2}+\tfrac{i\pi}{2k})}{\sinh(\tfrac{\theta}{2}-\tfrac{i\pi}{2k})}\cdot
\frac{\cosh(\tfrac{\theta}{2}-\tfrac{i\pi}{2k})}{\cosh(\tfrac{\theta}{2}+\tfrac{i\pi}{2k})}\label{x11}\,.
}
Note that this supplies simple poles at $iu_{11}^2=\frac{i\pi}k$ corresponding to the direct channel bound state $\langle 1,0\rangle$, and at $i(\pi-u_{11}^2)=i\pi-\frac{i\pi}k$, corresponding to the cross channel bound state $\langle 1,0\rangle$. We remark that it is very common in the construction of integrable S-matrix theories to take a tensor product
structure of the form \eqref{smfull}  \cite{Hollowood:1993fj}. The resulting S-matrix are trigonometric generalizations of the S-matrices of the principal chiral models which are obtained in the rational limit $q\to1$. 

By applying the bootstrap equations it follows that $X_{ab}(\theta)$ has four simple poles at $iu_{ab}^{a+b}$, $iu_{ab}^{|a-b|}$ and their crossed positions $i(\pi-u_{ab}^{a+b})$ and $i(\pi-u_{ab}^{|a-b|})$. If we define the standard S-matrix building blocks
\EQ{
[x]=\{x\}\{2k-x\}\ ,\qquad\{x\}=(x-1)(x+1)\ ,\qquad
(x)=\frac{\sinh(\frac\theta2+\frac{i\pi x}{4k})}
{\sinh(\frac\theta2-\frac{i\pi x}{4k})}\ ,
}
then
\EQ{
X_{ab}(\theta)=[a+b-1][a+b-3]\cdots[|a-b|+1]\ .
\label{smb}
}

\subsection{The closure of the bootstrap}

It is clear from the mass formula \eqref{mass} that the particle states can only exist for $a<2k$. Indeed, the bound state pole in $S_{ab}(\theta)$ at $\theta=\frac{i\pi(a+b)}{2k}$ moves off the physical strip for $a+b>2k$. So the spectrum of states must be bounded. The situation is very similar to the breather states in the sine-Gordon theory \cite{Zamolodchikov:1978xm}. Their masses are also given by a formula similar to \eqref{mass} with
\EQ{
k=\frac{8\pi}{\beta^2}\left(1-\frac{\beta^2}{8\pi}\right)\ ,
}
where $\beta$ is the sine-Gordon coupling. In the sine-Gordon case, the breather spectrum is actually cut off at $a\sim k$. The potential bound state pole in $S_{ab}(\theta)$ for $a+b>k$, but $<2k$, so that it is still on the physical strip, is actually an anomalous threshold arising from a graph involving the soliton states of the theory and is therefore not a bound state pole.\footnote{A careful explanation of this appears in \cite{Dorey:1996gd}.}

Since the mechanism above is not available for the present theories, there are two other ways that the additional poles may be removed. The first is inspired by the S-matrix of the non-simply-laced Toda theories \cite{Delius:1991kt,Corrigan:1993xh}. The theories associated to the pair of affine Lie algebras $(c_n^{(1)},d_{n+1}^{(2)})$ have a mass spectrum of the form \eqref{mass} with $H=2k$ and $a=1,2,\ldots,n$, where $n$ is the largest integer $\leq k$. The 2-point couplings are precisely those of figure~\ref{f8}  but with $a,b,a+b$ all restricted to $\leq n$. The S-matrix elements can be written as in \eqref{smb} above with a modified building block
\EQ{
\{x\}=\frac{(x-1)(x+1)}{(x+B-1)(x-B+1)}\ ,\qquad B=2(k-n)\ ,
}
where $0\leq B\leq 2$. In the Toda theory, $B$ is determined by the Toda coupling as
\EQ{
B=\frac1{2\pi}\cdot\frac{\beta^2}{1+\beta^2/4\pi}\ .
}
In the modified S-matrix factor $X_{ab}(\theta)$, the simple pole of the original S-matrix at $\theta=\frac{i\pi(a+b)}{2k}$, for $a+b>k$, is now absent. There are new simple poles on the physical strip whose origin is not due to bound states but in a generalized Coleman-Thun mechanism. Notice, however, that the resulting S-matrix is not an analytic function  of $k$.

Another way in which the spectrum can truncate with the additional poles on the physical strip for $a+b>k$ being removed, happens when $q$ is a root of unity. Although the representation theory of the quantum group $U_q(\mg)$ has not been developed in this case, we can infer what will happen 
by considering the behaviour of the bosonic subalgebra $U_q(\msl(2))\times U_q(\msl(2))$. In this case, it is well known that when $q=e^{i\pi/k}$ with $k$ a positive integer, then the series of representations of spin $\tfrac m2$ are truncated to the finite set $m=0,1,\ldots,k-2$. Since the representations $\langle a-1,0\rangle$ of $U_q(\mg)$ contain the $U_q(\mg_0)$ representations in \eqref{juu}, i.e.~$(a,0)\oplus(a-1,1)\oplus(a-2,0)$, 
this implies that the spectrum of states only includes the finite set $a=1,2,\ldots,k$. 
Representations near the top of the tower with $a=k$ and $k-1$ have a modified $U_q(\mg_0)$ content. For example, in the truncated representation theory, the representation $a=k$, that is $\langle k-1,0\rangle$, consists of the $U_q(\mg_0)$ representation $(k-2,0)$ only, while for $a=k-1$ we have $\langle k-2,0\rangle=(k-2,1)\oplus(k-3,0)$.
The truncated spectrum of states matches the semi-classical spectrum of soliton states in the SSSSG theories exactly \cite{Hollowood:2011fm,Hollowood:2011fq}. Notice that the S-matrix theory actually only makes sense for $k>2$ in order that the fundamental representation $\langle0,0\rangle$ exists. This suggests that $k$ in the S-matrix may be shifted relative to the $k$ in the action of the WZW model by a finite amount, namely $k\to k+2$, which is a common feature of the quantization of a WZW model, where 2 is the dual Coxeter number of $\msl(2)$. However, in the perturbative computation \cite{Hoare:2011fj} and in the related $\mathcal{N}=(2,2)$ supersymmetric sine-Gordon theory there is no evidence of such a shift. These issues will require further analysis to reconcile. 

\section{Discussion}

In this paper, we have shown how a conventional relativistic factorizable S-matrix can be constructed in an algebraic setting that is continuously connected to the non-relativistic S-matrix that describes the magnons on the string world sheet in $\text{AdS}_5\times S^5$. It would be interesting to see whether there exists a consistent, but necessarily non-relativistic, S-matrix theory that interpolates between our S-matrix and the magnon S-matrix. This would provide a quantum version of the classical picture of a Hamiltonian structure that interpolates the string world-sheet theory and the relativistic semi-symmetric space sine-Gordon theory.

We have not completed the analysis of the full S-matrix and shown that all the singularities on the physical strip can be accounted for as bound-state poles or anomalous thresholds and that the bootstrap closes: this remains for future work. However, the picture we have arrived at is very compelling and involves a delicate meshing of the representation theory of the affine quantum supergroup and the bootstrap/fusion procedure of S-matrix theory. 

The algebraic setting lying behind the family of S-matrix theories is rich and fascinating 
and at its heart seems to be a quantum deformation of a centrally extended loop superalgebra $\msl(2|2)^{(\sigma)}$. The R-matrices are also associated to other integrable systems, namely the one-dimensional Hubbard model and its deformations \cite{Beisert:2008tw}.
In a recent paper \cite{Beisert:2011wq} the full R-matrix  was shown to be related to what appears to be a different affine extension than our $\msl(2|2)^{(\sigma)}$, and it would be useful to gain an overview of how all the different facets of $\mpsl(2|2)\ltimes\mathbb R^3$ and its affinizations and $q$-deformations are related.

In the context of the magnon S-matrix, one notable feature  has been the apparent inability to use the bootstrap procedure \cite{Chen:2006gq, Roiban:2006gs, Chen:2006gp, Dorey:2007xn, Dorey:2007an,Arutyunov:2008zt}. It would clearly be interesting to re-visit this issue given that we have found that for the relativistic S-matrix the bootstrap equations seem to mesh perfectly with the peculiar representation theory of a $q$-deformed Lie superalgebra.

\section*{Acknowledgements}

We would like to thank Arkady Tseytlin for helpful comments on a draft of this paper.

\noindent
BH is supported by EPSRC.

\noindent
JLM thanks LPTHE~(UPMC Paris--CNRS) for the kind hospitality while this work was in progress.
He also acknowledges the support of MICINN (FPA2008-01838 and 
FPA2008-01177), Xunta de Galicia (Consejer\'\i a de Educaci\'on and INCITE09.296.035PR), the 
Spanish Consolider-Ingenio 2010
Programme CPAN (CSD2007-00042), and FEDER.

\startappendix

\Appendix{The Co-Product} \label{A}

For the relativistic Pohlmeyer-reduced theory the interpretation of the triply extended algebra as a finite subalgebra of an affine superalgebra $\mathfrak{sl}(2|2)^{(\s)}$ (discussed in section \ref{s3.1}) naturally gives the reality conditions given by \eqref{rcpr}. These differ from the ones that are usually used in the context of the $\text{AdS}_5 \times S^5$ string theory/${\cal N }= 4$ SYM magnon discussion~\eqref{rcgs}.

The reality conditions \eqref{rcgs} are related to those in \eqref{rcpr} with $\vpt=\frac{\pi}{2}$ by a change in the basis of the generators. This is explicitly given by
\ALAT{2}{
\notag &\tilde{\mR}^a{}_b = \mR^a{}_b\ , \qquad & &\tilde{\mL}^\alpha{}_\beta  = \mL^\alpha{}_\beta\ ,
\\\notag&\tilde{\mQ}^\alpha{}_b = \frac{1}{\sqrt{2}}(\mQ^\alpha{}_b -
\varepsilon^{\alpha\beta}\varepsilon_{ba}\mS^a{}_\beta)\ , \qquad&
&\tilde{\mS}^a{}_\beta  = \frac{1}{\sqrt{2}}(\mS^a{}_\beta +
\varepsilon^{ab}\varepsilon_{\beta\alpha}\mQ^\alpha{}_b)\ ,
\\&\tilde{\mP}  = \frac{1}{2}(\mP + \mK) - \mC \ , &
&\tilde{\mK}  = \frac{1}{2}(\mP + \mK) + \mC \ ,\label{cob}
\\\notag&&&\hspace{-2.5cm}\tilde{\mC}  = \frac{1}{2}(\mP - \mK)\ .
}

Moreover, this change of basis preserves the commutation relations \eqref{com1}, \eqref{com2}, i.e. it is an element of the outer-automorphism group $SL(2,\C)$. In particular it is in the $SL(2,\R)$ subgroup that contains those automorphisms that amount to a change of the basis for the generators of the real algebra.

For convenience, in this appendix we denote the eigenvalues of the central charges as $(P,\,K,\,C)$ and $(\tilde P,\,\tilde K,\,\tilde C)$ for the two different bases. Further we will refer to the basis satisfying the hermiticity relations \eqref{rcpr}  with $\vpt=\frac{\pi}{2}$ as basis (i) and to that satisfying \eqref{rcgs} as basis (ii).

For the soliton representation the central charges acting on the one-particle state  have the eigenvalues
\ALAT{3}{
\notag P=&-[\tfrac12]_q e^{-\o} \ , \qquad & 
K=& [\tfrac12]_q e^\o \ , \qquad &
C=&0 \ ,
\\
\tilde P = &[\tfrac12]_q \sinh \o \ , \qquad &
\tilde K = &[\tfrac12]_q \sinh \o \ , \qquad &
\tilde C = &[\tfrac12]_q \cosh \o \ .\label{prevtilde}
}
Of particular interest is the final equation where we see that the eigenvalue for the central charge $\tilde \mC$ is proportional to the two-dimensional energy. This is reminiscent of the magnon representation for which this third central extension is identified with the Hamiltonian.

It is interesting to note that for the relativistic representation in either basis the Lorentz symmetry can be identified with the subgroup of the real outer-automorphism group $SL(2,\R)$ that preserves the hermiticity relations. (The full subgroup of the complex outer-automorphism group $SL(2,\C)$ that preserves the hermiticity relations is $SU(1,1)$ \cite{Arutyunov:2009ga}.)

{}From \eqref{prevtilde} it is apparent that the Lorentz symmetry grading is clearer in basis (i) with the eigenvalue for the central charge $\mC$ vanishing. One may wonder if this change of basis could allow one to identify a hidden Lorentz symmetry of the magnon representation. To proceed we quote the central charge eigenvalues for the magnon representation \cite{Beisert:2005tm,Beisert:2006qh,Arutyunov:2006yd}
\ALAT{3}{
\tilde P &= -ge^{i(\frac{p}{2}+2\xi)}\sin\frac{p}{2} \ , \qquad &
\tilde K &= -ge^{-i(\frac{p}{2}+2\xi)}\sin\frac{p}{2} \ , \qquad &
\tilde C^2 &= \frac{1}{4}+g^2\sin^2\frac{p}{2}\ ,
}
where $p$ is the world sheet momentum, and identifying $\tilde C$ with the Hamiltonian the final equation gives the dispersion relation. $\xi$ is an arbitrary phase. The usual choice that is made is $\xi = 0$ for which the coproduct is given by \eqref{coprod} with $q=1$ and $\mU = e^{i \bP}$ where $\bP$ is the world sheet momentum operator. One can choose different values for $\xi$, however the coproduct needs to be modified accordingly \cite{Arutyunov:2009ga}.

Transforming into the alternative basis \eqref{cob}  we find 
\EQ{
P &= -g \sin\frac{p}{2}\cos\Big(\frac{p}{2}+2\xi\Big) +
\tilde C(p) \ , \\ 
K &=-g \sin\frac{p}{2}\cos\Big(\frac{p}{2}+2\xi\Big)
 - \tilde C(p) \ , \\ 
 C &= ig \sin\frac{p}{2}\sin\Big(\frac{p}{2}+2\xi\Big)\ .
}
Choosing $\xi=-\frac{p}{4}$ (recall that one needs to modify the coproduct accordingly) the central charge $C$ vanishes as in the relativistic Pohlmeyer reduced theory.

In this basis, and with this choice for $\xi$, one may expect to be able to associate a Lorentz symmetry grading to the generators. Indeed for the generators acting on the single-particle states this is just
\ALAT{1}{
P & \rightarrow e^{-\lambda} P \ ,\qquad\qquad  K  \rightarrow e^{\lambda} K \ , \notag
\\\sin \frac{p}{2} &\rightarrow -\frac{e^{-\lambda} P +e^{\lambda}K}{2g} =  \sin \frac{p}{2}\cosh \lambda  +\tilde C(p)\sinh \lambda  \ .\label{gslorentz}
}
While this transformation preserves the reality of $P$ and $K$ it clearly does not preserve the reality of the world sheet momentum $p$. 

Even so we may ask if it describes some formal hidden Lorentz symmetry of the Green-Schwarz theory. However, this appears not to be case as the symmetry does not extend to the action of the symmetry on tensor product representations. In particular this is a consequence of the coproduct \eqref{coprod} being constructed with the generators satisfying the reality conditions \eqref{rcgs}. By constructing the coproduct for the alternative basis using \eqref{cob} one can see explicitly that the symmetry identified in \eqref{gslorentz} does not extend to higher representations.

\end{document}